\documentstyle{report}
\catcode`@=11 
\def\thechapterhead{\Roman{chapter}.}
\def\thesectionhead{\Alph{section}.}

\def\@sect#1#2#3#4#5#6[#7]#8{\ifnum #2>\c@secnumdepth
     \def\@svsec{}\else                                           
     \refstepcounter{#1}\edef\@svsec{\csname the#1head\endcsname\hskip 1em }\fi
     \@tempskipa #5\relax
      \ifdim \@tempskipa>\z@ 
        \begingroup #6\relax
          \@hangfrom{\hskip #3\relax\@svsec}{\interlinepenalty \@M #8\par}
        \endgroup
       \csname #1mark\endcsname{#7}\addcontentsline
         {toc}{#1}{\ifnum #2>\c@secnumdepth \else
                      \protect\numberline{\csname the#1head\endcsname}\fi
                    #7}\else
        \def\@svsechd{#6\hskip #3\@svsec #8\csname #1mark\endcsname
                      {#7}\addcontentsline               
                           {toc}{#1}{\ifnum #2>\c@secnumdepth \else
                             \protect\numberline{\csname the#1head\endcsname}\fi
                       #7}}\fi
     \@xsect{#5}}

\def\appendixname{APPENDIX}     %GMD <---------
\def\appendix{\par
 \setcounter{chapter}{0}
 \setcounter{section}{0}
 \def\@chapapp{\appendixname} %GMD <----------
 \def\thechapter{\Alph{chapter}}
 \def\thechapterhead{\Alph{chapter}}
 \def\thesection{\thechapter.\arabic{section}}
 \def\thesectionhead{\thechapterhead.\arabic{section}}}
\def\@makechapterhead#1{ {\setbox0=\hbox{#1} \parindent 0pt 
 \begin{raggedright} 
 \Large \bf 
 \ifnum \c@secnumdepth >\m@ne \ifdim \wd0=0pt \@chapapp{} \thechapterhead
   \else \thechapterhead \hspace{1ex} \ #1\fi \end{raggedright}\par 
 \fi \nobreak \vskip 2ex }}
\def\@makeschapterhead#1{ { \parindent 0pt \begin{raggedright}
 \Large \bf  #1 \end{raggedright}\par 
 \nobreak \vskip 2ex } }
\def\thebibliography#1{\chapter*{\bibname\@mkboth 
 {\uppercase{\bibname}}{\uppercase{\bibname}}}\list 
 {[\arabic{enumi}]}{\settowidth\labelwidth{[#1]}\leftmargin\labelwidth 
 \advance\leftmargin\labelsep 
 \usecounter{enumi}} 
 \def\newblock{\hskip .11em plus .33em minus -.07em} 
 \sloppy 
 \sfcode`\.=1000\relax} 
\def\bibname{REFERENCES} 
\def\chapter{\par \vspace{3.5ex} \thispagestyle{plain} \global\@topnum\z@
\@afterindentfalse \secdef\@chapter\@schapter}
\def\section{\@startsection {section}{1}{\z@}{-2.0ex plus -1ex minus
 -.2ex}{1.0ex plus .2ex}{\Large\bf}}
\def\subsection{\@startsection{subsection}{2}{\z@}{-2.0ex plus -1ex minus
 -.2ex}{0.5ex plus .2ex}{\large\bf}}
\def\citenum{\@ifnextchar [{\@tempswatrue\@zcitex}{\@tempswafalse\@zcitex[]}}
\def\@zcitex[#1]#2{\if@filesw\immediate\write\@auxout{\string\citation{#2}}\fi
  \def\@citea{}\@zcite{\@for\@citeb:=#2\do
    {\@citea\def\@citea{,\penalty\@m\ }\@ifundefined
       {b@\@citeb}{{\bf ?}\@warning
       {Citation `\@citeb' on page \thepage \space undefined}}%
\hbox{\csname b@\@citeb\endcsname}}}{#1}}
\def\@zcite#1#2{{#1\if@tempswa , #2\fi}}

\def\citevb{\@ifnextchar [{\@tempswatrue\@citexvb}{\@tempswafalse\@citexvb[]}}
\def\@citexvb[#1]#2{\if@filesw\immediate\write\@auxout{\string\citation{#2}}\fi
  \def\@citea{}\@cite{\@for\@citeb:=#2\do
    {\@citea\def\@citea{--\penalty\@m}\@ifundefined
       {b@\@citeb}{{\bf ? \@citeb}\@warning
       {Citation `\@citeb' on page \thepage \space undefined}}%
\hbox{\csname b@\@citeb\endcsname}}}{#1}}
\newbox\uphookbox
\setbox\uphookbox=\hbox{\vrule width 1.2pt depth 3.0pt height 1.18pt}
\dp\uphookbox=0\p@\def\uphook{\copy\uphookbox}

\def\uphookbracefill#1#2{%
\setbox1=\hbox{$\m@th\displaystyle#1$\hskip-1.2pt}%
\setbox2=\hbox{$\m@th\displaystyle#2$\hskip-1.2pt}%
$\m@th\hskip0.5\wd1\uphook\leaders\vrule\hfill\uphook\hskip0.5\wd2$}
\def\uphookbrace#1#2#3{\mathop
   {\vbox
      {\ialign{##\crcr
         \noalign{\kern3\p@}%
         \uphookbracefill{#1}{#3}\crcr
         \noalign{\kern3\p@\nointerlineskip}%
         $\hfil\displaystyle{#1#2#3}\hfil$\crcr}
      }%
   }%
\limits}
\def\sqr#1#2{{\vcenter{\hrule height.#2pt
\hbox{\vrule width.#2pt height#1pt \kern#1pt
\vrule width.#2pt}
\hrule height.#2pt}}}

\newif\ifseparateabstract\separateabstractfalse
\def\abstract#1{\gdef\@abstract{#1}}
\def\pacs#1{\gdef\@pacs{#1}}
\def\maketitle{\begin{titlepage}%
 \let\footnotesize\small      % Local definitions to make \thanks produce
 \setcounter{page}{0}%
 \null
 \vfil
 \vskip 30pt                  % To adjust centering.
 \vbox                                            
  {\begin{center}    
   {\Large\bf \@title \par}        % Set title in \Large\bf
   \vskip 3em                  % Vertical space after title.
   {                    
     \lineskip .75em
     \begin{tabular}[t]{c}\@author 
     \end{tabular}\par}      
    \vskip 1.5em               % Vertical space after author.
   { \@date \par}        
  \end{center} \par}\nopagebreak % end of the vbox
  \@thanks                                                              
 \setcounter{footnote}{0}       % Footnotes start at zero again.
 \ifseparateabstract 
   \vfil\null                                                
   \end{titlepage}   
   \begin{titlepage}
   \null\vfil
 \else
   \vskip 0pt plus 0.6fil\penalty5000
 \fi
 \vbox
   {\begin{center}                                                  
     {\large\bf Abstract}       % Abstract heading set in \large\bf. 
   \end{center}\nopagebreak
   \@abstract    
   \par\vskip4ex \noindent\@pacs} 
 \vfil\null\end{titlepage}
\let\thanks\relax
\gdef\@thanks{}\gdef\@author{}\gdef\@title{}\let\maketitle\relax
\gdef\@abstract{}\gdef\@pacs{}}
\gdef\@thanks{}\gdef\@author{}\gdef\@title{}
\gdef\@abstract{}\gdef\@pacs{}
\catcode`@=12 
\newcommand{\tr}{\mathop{\rm tr} \nolimits}
\newcommand{\Tr}{\mathop{\rm Tr} \nolimits}

\newcommand{\trg}{\mathop{\rm trg} \nolimits}

\newcommand{\asymx}{\mathop{\sim}}
\newcommand{\asym}[1]{\mathrel{\asymx_{#1}}}

\newcommand{\half}{{\textstyle \frac{1}{2}}}
 
\newdimen\savebaselineskip 
\savebaselineskip=\baselineskip 

\catcode`@=11
\def\lsim{\mathrel{\mathpalette\@versim<}}
\def\gsim{\mathrel{\mathpalette\@versim>}}
\def\@versim#1#2{\vcenter{\offinterlineskip
        \ialign{$\m@th#1\hfil##\hfil$\crcr#2\crcr\sim\crcr } }}
\def\ltgt{\mathrel{\mathpalette\@vergt<}}
\def\gtlt{\mathrel{\mathpalette\@verlt>}}
\def\@vergt#1#2{\vcenter{\offinterlineskip
        \ialign{$\m@th#1\hfil##\hfil$\crcr#2\crcr>\crcr } }}
\def\@verlt#1#2{\vcenter{\offinterlineskip
        \ialign{$\m@th#1\hfil##\hfil$\crcr#2\crcr<\crcr } }}
\catcode`@=12
\newcommand{\opfone}{\bf 1}

\newcommand{\diag}{\mathop{\rm diag} \nolimits}

\newcommand{\Trg}{\mathop{\rm Trg}}
\renewcommand{\Tr}{\mathop{\rm Tr}}
\newcommand{\Detg}{\mathop{\rm Detg} \nolimits}
\newcommand{\detg}{\mathop{\rm detg} \nolimits}
\newcommand{\re}{\mathop{\rm Re} \nolimits}
\newcommand{\im}{\mathop{\rm Im} \nolimits}

\begin{document}
\title{Introduction to the Supersymmetry Method for the Gaussian 
       Random-Matrix Ensembles}
\author{
{\bf Josef~A.~Zuk}\thanks{Current address: Aeronautical Research
     Laboratory, Air Operations Division, P.O.~Box 4331, Melbourne, 
     Australia. (E-mail: Josef.Zuk@dsto.defence.gov.au)} \\
     {\em Max--Planck--Institut f\"ur Kernphysik, D-69029 Heidelberg, 
     Germany} \\
     \& \\
     {\em Department of Physics, University of Manitoba,
     Winnipeg, R3T 2N2, Canada} 
       }
\date{}
%     \hfill November 1994}
\abstract{
This article is intended to provide a pedagogical introduction to the 
supersymmetry method for 
performing ensemble-averaging in Gaussian random-matrix theory. 
The method is illustrated by a detailed calculation of the simplest 
non-trivial physical quantity, namely, the second-order correlation in the
density of states for two different energies within the spectrum (commonly
known as the density-density correlator) for a system described by a
random Hamiltonian matrix belonging to the Gaussian unitary ensemble. 
}
\maketitle 

\section*{CONTENTS}
\begin{tabbing}
III.XXX \= A.XX \= Hubbard-Stratonovich Transformation \kill
I.      \> Introduction \\
II.     \> Density of States \\
III.    \> Superalgebra \\
IV.     \> Supersymmetry Formalism \\
        \> A.   \> Basic Ideas \\
        \> B.   \> Generating Function \\
        \> C.   \> Ensemble Average \\
        \> D.   \> Hubbard-Stratonovich Transformation \\
        \> E.   \> Integration Supermanifold \\
        \> F.   \> Convergence \\
        \> G.   \> Saddle-Point Equation \\
        \> H.   \> Decoupling of Massive Modes \\
V.      \> Superintegration \\
        \> A.   \> Coset Parametrization \\
        \> B.   \> The Measure \\
Acknowledgements \\
Appendix A \\
Appendix B \\
Appendix C \\
References 
\end{tabbing}
\newpage

\chapter{Introduction}
In many physical problems, especially in nuclear physics and condensed-matter 
physics, the quantum mechanical Hamiltonian, which models the underlying 
physical complexity, involves a random matrix belonging to one of the three 
standard Gaussian ensembles: the Gaussian unitary ensemble (GUE), the 
Gaussian orthogonal ensemble (GOE) or the Gaussian symplectic ensemble (GSE). 
The appropriate choice is dictated by the symmetries of the theory 
\cite{Meh91}. 

In condensed-matter physics, such problems arise in the study of electron 
localization phenomena in disordered conductors and semi-conductors. 
In the mesoscopic regime, applications include universal conductance 
fluctuations \cite{IWZ}, Aronov-Altshuler-Spivak oscillations \cite{JZ,AMG}
and persistent currents \cite{AIMW}. 
These are all weak-localization effects. 
In the strongly localized domain, problems such as long disordered quantum 
wires \cite{MMZ} and the integer quantum Hall effect \cite{WZ}
are amenable to a random-matrix formulation. 
In the ballistic regime, where disorder is negligible, classical chaos in 
the dynamics of electron scattering gives rise to the complexity that can 
be described within random-matrix theory. 
Thus, another recent important application has been the study of electron 
transport across microstructures, constructed in the shape of classically 
chaotic billiards, in the presence of a variable external magnetic field. 
This allows one to investigate the continuous crossover between two 
random-matrix ensembles (GOE~$\to$~GUE) as probed by the mean conductance 
and its correlations for increasing magnetic field \cite{PWZL}. 
Also, connections between the Sutherland-Calogero model and the random-matrix 
ensembles have been established and discussed in the literature 
\cite{WD,SLA}.
A brief review of applications to nuclear physics can be found in 
Ref.~\citenum{BW}.  
Finally, the latest application of the supersymmetry techniques described
here has emerged in elementary particle physics, where random-matrix
theory has been used to study chiral symmetry breaking in QCD \cite{QCD}. 

One is often interested in computing ensemble averages of physical quantities 
involving traces of products or products of traces of resolvents of the 
Hamiltonian. Such quantities include the density of states and its 
correlation functions, products of S-matrix elements, as well as the 
conductance (by virtue of being itself a sum over products of S-matrix
elements, as can be seen from the Landauer formula, which states that 
\mbox{$G = (e^2/h)\tr t^\dagger t$} 
with the matrix $t$ denoting the transmission part of the S-matrix) 
and its higher moments and auto-correlations. 
In this situation, it is convenient to proceed by first constructing a 
generating function for resolvents and expressing the desired quantities as 
derivatives of the generating function with respect to the source. 
The supersymmetry method \cite{Efetov,VWZ}
provides a mathematical tool for performing the 
ensemble average of the entire generating function, thereby avoiding the 
necessity for expanding in powers of the random matrix, and subsequently 
resumming after ensemble averaging (as is done in impurity perturbation 
theory, for example). Therefore, it is a non-perturbative approach which 
can provide results beyond the domain of validity of expansion techniques. 
It is also a useful alternative to the method of 
orthogonal polynomials in many cases \cite{Meh91}. 
Indeed, where coupling to external channels is involved, such as in the 
calculation of S-matrix correlations \cite{IWZ}, orthogonal polynomials 
have not been applied and may not be amenable to such problems. 

Furthermore, non-perturbative results are especially useful whenever a
small number of external channels, $M$, is coupled to the random system,
because the perturbation expansion here proceeds essentially in powers of
$1/M$. This situation can typically occur in the problem of
compound-nucleus scattering with few open decay channels \cite{VWZ}, or in
the conductance problem for electronic microstructures connected to few
electron-mode leads \cite{PWZL}. An extreme case is provided by isolated 
(\mbox{$M=0$})
non-dissipative systems in dealing with the so-called zero mode. The
problem of persistent currents gives one such example \cite{AIMW}. 

The foregoing remarks have hopefully illustrated the fact 
that random-matrix theory is a subject of much current interest, and the 
supersymmetric approach is playing an increasingly important role in it
--- its use having been very rapidly increasing over the past few years. 
It is difficult, however, for the novice to easily grasp the elements of
the supersymmetry method simply by studying the topical literature. 
Applications to real physics problems typically involve calculations
rendered long and laborious by the presence of parameters inducing
explicit symmetry breaking, couplings to external degrees of freedom, the
necessity for complicated source terms, continuous spatial dimensions, and
the large dimensionality of the underlying matrix spaces required for
evaluation of higher-order correlators. 
Usually also, other extraneous ingredients come into play in a given
physical problem, such as statistical scattering theory. All these
complications tend to obscure the essentials of the supersymmetry method.
The purpose of the present article is to bridge the gap by providing a
primer for more realistic calculations. 

To this end, 
we consider here a quantum mechanical theory whose Hamiltonian is taken to 
be purely 
a random $N\times N$ Hermitian matrix belonging to the Gaussian unitary 
ensemble (GUE). 
By definition, a random matrix $H$ is said to belong to the GUE if (i) its
diagonal elements $H_{\mu\mu}$ and the real and imaginary parts of its
off-diagonal elements $H_{\mu\nu}$, for 
\mbox{$\mu < \nu$}, 
are statistically independent, and (ii) the probability 
\mbox{$P(H){\cal D}H$} 
that the system belongs to the volume element 
\begin{equation} 
{\cal D}H = \prod_{\mu=1}^N dH_{\mu\mu} \cdot \prod_{\mu < \nu} 
     d(\re H_{\mu\nu}) d(\im H_{\mu\nu}) \;,  
\label{vol-el}
\end{equation} 
is invariant under every automorphism 
\begin{equation} 
H \mapsto H' = U^{-1}HU \;, 
\end{equation} 
where $U$ is any unitary matrix, i.e.\ 
\mbox{$P(H'){\cal D}H' = P(H){\cal D}H$}. 
It should be noted that the diagonal elements $H_{\mu\mu}$ are necessarily
real, as $H$ is Hermitian. 

Because of (i), $P(H)$ is a product of functions each of which depends on
a single variable (viz.\ the statistically independent elements 
$H_{\mu\mu}$, 
\mbox{$\re H_{\mu\nu}, \im H_{\mu\nu}, \mu<\nu$}). 
Then, as a consequence of the unitary invariance (ii), as originally
proven by Porter and Rosenzweig, it follows that the statistically
independent elements are Gaussian distributed with a common variance. A
derivation of this fact may be found in Section~2.6 of
Ref.~\citenum{Meh91}. 
Accordingly, the second moments of the matrix elements are given by 
\begin{equation} 
\overline{H_{\mu\nu}H_{\mu'\nu'}} = \frac{\lambda^2}{N} 
     \delta_{\mu\nu'}\delta_{\nu\mu'} \;, 
\label{moments}
\end{equation} 
and we take them to have zero mean 
\mbox{$\overline{H_{\mu\nu}} = 0$}. 
The appearance of the matrix dimension $N$ in Eq.~(\ref{moments}) provides
a convenient normalization for the common strength $\lambda$, while the
structure of the Kronecker deltas reflects the required statistical
independence as well as the Hermiticity of $H$.
For obvious reasons, we refer to 
\mbox{$\mu,\nu$} 
as the level indices. 
The overbar represents ensemble averaging; and since we assume a Gaussian
distribution of Hamiltonian matrices, the mean and second moments (as 
specified above) serve to define it uniquely. 

The ensemble average of any function 
\mbox{$f(H)$} 
of the random matrix $H$ can be represented as 
\begin{equation} 
\overline{f(H)} = \frac{1}{{\cal N}}
     \int {\cal D}H\, f(H) \exp\Bigl\{-\frac{N}{2\lambda^2} 
     \Tr H^2\Bigr\} \;, 
\label{avg-f}
\end{equation} 
with the measure 
\mbox{${\cal D}H$} as given in Eq.~(\ref{vol-el})
and the constant ${\cal N}$ chosen to ensure unit normalization. 
Now let us suppose that the function
\mbox{$f(H)$}, 
whose average we are considering in Eq.~(\ref{avg-f}), depends only on the
eigenvalues of $H$, viz., the energy levels 
\mbox{$E_1,E_2,\ldots,E_N$}. 
This will happen, for example, if 
\mbox{$f(H)$} 
satisfies the symmetry property 
\begin{equation} 
f(U^{-1}HU) = f(H) 
\end{equation} 
for any unitary matrix 
\mbox{$U \in {\rm U}(N)$}. 
To see this, let us note that any Hermitian matrix $H$ can always be
diagonalized by an 
\mbox{$N \times N$} 
unitary matrix $U_H$ from the coset space 
\mbox{${\rm U}(N)/[{\rm U}(1)]^N$}, 
i.e.
\begin{equation} 
H = U_H H_{\rm D} U^{-1}_H \;, 
\label{diag-h}
\end{equation} 
where $H_{\rm D}$ is the diagonal matrix of eigenvalues 
\mbox{$H_{\rm D} = \diag(E_1,E_2,\ldots,E_N)$}. 
Then we have 
\begin{eqnarray}
f(H) & = & f(U^{-1}_H H U_H) 
\nonumber \\
& = & f(\diag(E_1,E_2,\ldots,E_N)) 
\nonumber \\
& \equiv & f(E_1,E_2,\ldots,E_N) \;. 
\end{eqnarray}
Moreover, because of the decomposition (\ref{diag-h}), we can make the
change of integration variables in Eq.~(\ref{avg-f}) from the matrix
elements 
\mbox{$H_{\mu\nu}$} 
to the $N$ eigenvalues 
\mbox{$E_1,E_2,\ldots,E_N$} 
and 
\mbox{$N^2-N$} 
real angles that are needed to parametrize the space of unitary matrices
$U_H$. 

This allows us to explicitly perform the 
\mbox{$N^2-N$} 
angular integrations implicit in Eq.~(\ref{avg-f}), and thereby express it
as 
\begin{equation} 
\overline{f(E_1,\ldots,E_N)} = \frac{V_N}{{\cal N}}\int 
     \prod_{i=1}^N dE_i\, J(E_1,\ldots,E_N) f(E_1,\ldots,E_N) 
     \exp\Bigl\{-\frac{N}{2\lambda^2}\sum_{i=1}^N E_i^2\Bigr\} \;, 
\label{star}
\end{equation} 
where 
\mbox{$J(E_1,\ldots,E_N)$} 
denotes the Jacobian for the transformation of integration variables, and 
\begin{equation} 
V_N \equiv \int d\Omega_N(U) \;, 
\end{equation} 
being the integral of the Haar measure 
\mbox{$d\Omega_N(U)$} 
over the coset space 
\mbox{${\rm U}(N)/[{\rm U}(1)]^N$}, 
expresses the volume of this space. 
If we proceed to write Eq.~(\ref{star}) in the form
\begin{equation} 
\overline{f(E_1,\ldots,E_N)} = \int\prod_{i=1}^N dE_i\, f(E_1,\ldots,E_N) 
     P(E_1,\ldots,E_N) \;, 
\end{equation} 
then we can immediately read off an expression for the joint probability
density function for the eigenvalues as being given by 
\begin{equation} 
P(E_1,\ldots,E_N) = \frac{V_N}{{\cal N}}J(E_1,\ldots,E_N) 
     \exp\Bigl\{-\frac{N}{2\lambda^2}\sum_{i=1}^N E_i^2\Bigr\} \;. 
\end{equation} 
To obtain the two-level distribution function, one integrates
over all eigenvalues $E_i$ except for $E_1$ and $E_2$: 
\begin{equation} 
p_2(E_1,E_2) =  \frac{V_N}{{\cal N}}\int 
     \prod_{i=3}^N dE_i\, J(E_1,E_2,\ldots,E_N)
     \exp\Bigl\{-\frac{N}{2\lambda^2}\sum_{i=1}^N E_i^2\Bigr\} \;. 
\label{p2}
\end{equation} 

To calculate the Jacobian $J$ \cite{PJFU}, let us begin by considering the
differential of Eq.~(\ref{diag-h}), viz.\
\mbox{$H = UH_{\rm D}U^{-1}$}, 
which yields 
\begin{equation} 
dH = dUH_{\rm D}U^{-1} + UdH_{\rm D}U^{-1} + UH_{\rm D}dU^{-1} \;. 
\end{equation} 
Consequently, 
\begin{equation} 
U^{-1}dHU = U^{-1}dUH_{\rm D} - H_{\rm D}U^{-1}dU + dH_{\rm D} \;, 
\label{hash} 
\end{equation} 
having used the relation 
\mbox{$dU^{-1}U = -U^{-1}dU$}. 
Now let 
\mbox{$(dH)$} 
denote the exterior product of the differentials of the independent 
elements of $H$, i.e. 
\begin{eqnarray} 
(dH) & = & \bigwedge_{\mu,\nu=1}^N dH_{\mu\nu} 
\nonumber \\ 
& = & \Bigl\{\bigwedge_{\mu=1}^N dH_{\mu\mu}\Bigr\} \wedge 
     \Bigl\{\bigwedge_{\mu<\nu} dH_{\mu\nu}\wedge dH^*_{\mu\nu} 
     \Bigr\} 
\nonumber \\ 
& = & (2i)^{N(N-1)/2}\Bigl\{\bigwedge_{\mu=1}^N dH_{\mu\mu}\Bigr\} \wedge 
     \Bigl\{\bigwedge_{\mu<\nu} d\re H_{\mu\nu}\wedge d\im H_{\mu\nu} 
     \Bigr\} \;, 
\label{wedge}
\end{eqnarray} 
given a suitable ordering of the matrix elements. From the linearity of
the wedge product, it follows that 
\begin{equation} 
(U^{-1}dHU) = q(U,U^*)(dH)
\label{one} 
\end{equation} 
for some polynomial 
\mbox{$q(U,U^*)$} 
of the independent matrix elements of $U$ and their complex conjugates,
which can be shown to have unit modulus 
\mbox{$|q(U,U^*)|=1$}. 

On the other hand, by expressing 
\mbox{$U = (\vec{u}_1,\vec{u}_2,\ldots,\vec{u}_N\,)$} 
in terms of column vectors $\vec{u}_i$, 
\mbox{$i = 1,2,\ldots,N$} 
(which correspond to the orthonormal eigenvectors of $H$), and then
substituting this form into Eq.~(\ref{hash}), one can easily show that the
exterior product of the RHS of Eq.~(\ref{hash}) simplifies to yield 
\begin{equation} 
(U^{-1}dHU) = \Bigl\{\prod_{1\le j<k\le N} (E_j-E_k)^2 \bigwedge_{i=1}^N 
     dE_i\Bigr\} \wedge (U^{-1}dU) \;. 
\label{two} 
\end{equation} 
After equating Eqs.~(\ref{one}) and (\ref{two}), and taking into account
Eq.~(\ref{wedge}), we can read off the volume element 
\begin{equation} 
{\cal D}H = d\Omega_N(U)J(E_1,\ldots,E_N) 
     \prod_{i=1}^N dE_i \;, 
\end{equation} 
with the Jacobian identified with 
\begin{equation} 
J(E_1,\ldots,E_N) = 2^{-N(N-1)/2}\prod_{1\le j<k\le N} (E_j-E_k)^2 
\end{equation} 
and 
\mbox{$d\Omega_N(U)$} 
denoting the Haar measure on the coset space 
\mbox{${\rm U}(N)/[{\rm U}(1)]^N$}.
Consequently, the joint probability density function assumes the form 
\begin{equation} 
P(E_1,\ldots,E_N) = C_N(\lambda)\prod_{1\le j<k\le N} (E_j-E_k)^2 
     \exp\Bigl\{-\frac{N}{2\lambda^2}\sum_{i=1}^N E_i^2\Bigr\} \;. 
\end{equation} 

The constant 
\mbox{$C_N(\lambda)$}, 
which is equal to 
\mbox{$C_N(\lambda) = 2^{-N(N-1)/2}{\cal N}^{-1}V_N$}, 
can be determined directly by explicitly calculating ${\cal N}$ and $V_N$.
On the other hand, it can also be fixed by using the fact that the joint
probability density function 
\mbox{$P(E_1,\ldots,E_N)$} 
integrated over all the eigenvalues $E_i$ must yield unity. Thus here, for
the sake of brevity, we shall take account of this fact and 
simply derive an expression for 
\mbox{$C_N(\lambda)$} 
by appealing to Mehta's integral \cite{Meh91}, which states that, for 
\mbox{$\beta = 1,2,4$}, 
\begin{equation} 
\prod_{i=1}^N\int_{-\infty}^{+\infty}dx_i\, 
     e^{-\half\sum_{i=1}^N x_i^2} \prod_{1\le j<k\le N} |E_j-E_k|^\beta 
     = (2\pi)^{N/2}\prod_{j=1}^N \frac{\Gamma(1+j\beta/2)} 
     {\Gamma(1+\beta/2)} \;. 
\end{equation} 
So, for the GUE value of 
\mbox{$\beta=2$}, 
we obtain 
\begin{eqnarray} 
\frac{1}{C_N(\lambda)} & = & \prod_{i=1}^N\int_{-\infty}^{+\infty} 
     dE_i\, e^{-N/2\lambda^2\sum_{i=1}^N E_i^2} 
     \prod_{1\le j<k\le N} (E_j-E_k)^2 
\nonumber \\ 
& = & \left(\frac{\lambda^2}{N}\right)^{N^2/2}(2\pi)^{N/2} 
     \prod_{j=1}^N j! \;. 
\end{eqnarray} 

The GUE is relevant for systems without invariance under time reversal.
Since our aim is to give a tutorial introduction to
the application of the supersymmetry method for random-matrix problems, we 
shall present a detailed step-by-step calculation of the simplest non-trivial
physical quantity in this theory, namely, the density-density correlator. 
Before doing so, however, we shall demonstrate that the second order
correlator of the density of states at two energies 
\mbox{$E_1,E_2$} 
within the spectrum of the theory coincides with the two-level distribution
function $p_2(E_1,E_2)$ that we have just discussed. 
One physical system to which our calculation applies directly is that 
composed of small disordered metallic grains that are subjected to an 
external magnetic field \cite{Efetov}.  

We shall begin our exposition by relating the connected part of the
density-density correlator 
\mbox{$C(E_1,E_2) = \left[\overline{\rho(E_1)\rho(E_2)}\right]_{\rm conn.}$}
to the ensemble average of the product of retarded and advanced Green's
functions 
\mbox{$G^\pm(E)$}, 
which are given as resolvents. 
We show that, in the large-$N$ limit, 
\begin{equation}
C(E_1,E_2) = \frac{1}{2\pi^2N^2}\re \left[\overline{G^+(E_1)G^-(E_2)} 
     \right]_{\rm conn.} \;.
\end{equation} 
Our discussion will then detail the construction of a generating function
$Z(\varepsilon)$, depending on source-term parameters 
\mbox{$\varepsilon = (\varepsilon_1,\varepsilon_2)$}, 
with respect to whose components twofold differentiation yields the
product of an advanced and a retarded Green's function. 
After having demonstrated how to perform the ensemble average exactly
with the aid of supersymmetry concepts, and how to extract the large-$N$
limit using saddle-point techniques, we shall arrive at the result: 
\begin{eqnarray} 
\overline{G^+(E_1)G^-(E_2)} & = & \left. \frac{1}{4} 
     \frac{\partial^2}{\partial\varepsilon_1\partial\varepsilon_2} 
     \overline{Z(\varepsilon)}\right|_{\varepsilon=0} 
\nonumber \\ 
& = & -\left(\frac{\pi}{2d}\right)^2\int{\cal D}Q\, \trg kQ_{11} \trg kQ_{22} 
     \exp\left\{\frac{i\pi\omega^+}{2d}\trg LQ\right\} \;,
\end{eqnarray} 
assuming 
\mbox{$E_1+E_2=0$} 
for simplicity, where 
\mbox{$\omega=E_1-E_2$}, 
\mbox{$\omega^+=\omega+i\epsilon$}, 
and with  
\mbox{$d=\pi\lambda/N$} 
denoting the mean level spacing at 
\mbox{$E=0$}. 
This is an integral over a coset manifold of four-dimensional
supermatrices $Q$, which we show how to parametrize and evaluate in order
to derive an expression for the connected part of the
density-density correlator that has been previously established in the
literature on random matrix theory by a variety of methods 
\cite{Meh91,Efetov,VZ}, viz., 
\begin{equation} 
C(\omega) =  \frac{1}{(\pi\lambda)^2}\biggl[ \pi\delta(x) -  
     \frac{\sin^2x}{x^2}\biggr] \;,
\end{equation} 
with 
\mbox{$x = \pi\omega/d$}. 
Since we assume 
\mbox{$E_1+E_2=0$}, 
we use the notation 
\mbox{$C(\omega) \equiv C(\omega/2,-\omega/2)$} 
for the functional dependence on the energy difference $\omega$. 

\chapter{Density of States} 
The density of states, normalized to unity, is given by 
\begin{equation} 
\rho(E) = \frac{1}{N}\sum_{n=1}^N \delta(E-E_n) 
        = \frac{1}{N}\Tr \delta(E-H) \;. 
\label{dos}
\end{equation} 
With this definition, it is clear that 
\begin{equation} 
\int_{-\infty}^{+\infty}dE\, \rho(E) = 1 \;. 
\label{unity}
\end{equation} 
Using the identity 
\begin{equation} 
\delta(E) = \frac{1}{2\pi i}\left[\frac{1}{E-i\epsilon} - 
     \frac{1}{E+i\epsilon}\right] \;, \quad \epsilon \to 0^+ \;, 
\end{equation} 
we obtain the representation
\begin{eqnarray} 
N\rho(E) & = & \frac{1}{2\pi i}\left[\Tr\frac{1}{E-H-i\epsilon} 
     - \Tr\frac{1}{E-H+i\epsilon}\right] 
\nonumber \\ 
& = & \frac{i}{2\pi}\left[G^+(E) - G^-(E)\right] 
\nonumber \\ 
& = & -\frac{1}{\pi}\im G^+(E) \;, 
\label{Gdiff}
\end{eqnarray} 
where 
\begin{equation} 
G^\pm(E) \equiv \Tr\frac{1}{E-H\pm i\epsilon} \;. 
\end{equation} 
We note that 
\mbox{$G^+(E)$} 
corresponds to a retarded Green's function, while 
\mbox{$G^-(E)$} 
is an advanced Green's function, i.e., if we introduce the time-dependent 
Green's functions 
\begin{equation} 
\tilde{G}^\pm(t) = \int_{-\infty}^{+\infty} \frac{dE}{2\pi}\, e^{-iEt} 
     G^\pm(E) \;, 
\end{equation} 
it follows that 
\begin{equation} 
\tilde{G}^\pm(t) = 0 \quad\mbox{for}\quad t \ltgt 0 \;. 
\end{equation} 

Given Eq.~(\ref{avg-f}) for the definition of the ensemble average and
Eq.~(\ref{dos}) for the density of states, we can write the
density-density correlator as 
\begin{eqnarray} 
\overline{\rho(E')\rho(E'')} & = & \frac{1}{{\cal N}N^2} \int {\cal D}H\, 
     \Tr\delta(E'-H)\Tr\delta(E''-H) e^{-N/2\lambda^2\Tr H^2} 
\nonumber \\ 
& = & \frac{1}{{\cal N}N^2}\sum_{j,k=1}^N\int\prod_{i=1}^N dE_i\, 
     J(E_1,\ldots,E_N)\int d\Omega_N(U)\, \delta(E'-E_j) 
     \delta(E''-E_k)e^{-N/2\lambda^2\sum_{i=1}^N E_i^2} 
\nonumber \\ 
& = & \frac{V_N}{{\cal N}}\int\prod_{i=3}^N dE_i\, 
     J(E',E'',E_3,\ldots,E_N)\exp\Bigl\{-\frac{N}{2\lambda^2}\Bigl( 
     {E'}^2+{E''}^2+\sum_{i=3}^N E_i^2\Bigr)\Bigr\} \;, 
\label{wiggle} 
\end{eqnarray} 
where 
\mbox{$J(E_1,\ldots,E_N)$} 
is the Jacobian for the change of variables from matrix elements of $H$ to
eigenvalues $E_i$ and orthonormal eigenvectors $\vec{u}_i$, as discussed
in the foregoing section; 
\mbox{$d\Omega_N(U)$} 
is the Haar measure over the space of unitary matrices formed from the
column eigenvectors 
\mbox{$U = (\vec{u}_1,\vec{u}_2,\ldots,\vec{u}_N\,)$}, 
and $V_N$ is the volume of this space. The final line in
Eq.~(\ref{wiggle}) follows from the fact that the Jacobian is totally
symmetric under all permutations of its arguments. Eq.~(\ref{wiggle}) can
be recast as 
\begin{equation} 
\overline{\rho(E_1)\rho(E_2)} = C_N(\lambda)\int\prod_{i=3}^N dE_i\, 
     J(E_1,E_2,\ldots,E_N)\exp\Bigl\{-\frac{N}{2\lambda^2}\sum_{i=1}^N 
     E_i^2 \Bigr\}\;,
\end{equation} 
from which it is evident, given Eq.~(\ref{p2}), that the density-density
correlator coincides with the two-level distribution function 
\mbox{$p_2(E_1,E_2)$}. 

We are interested in looking at the connected part of the ensemble-averaged 
density-density correlator, 
\begin{equation} 
C(E_1,E_2) = \overline{\rho(E_1)\rho(E_2)} - \overline{\rho(E_1)} \cdot 
     \overline{\rho(E_2)} \;. 
\end{equation} 
Using the representation (\ref{Gdiff}), and the fact that for large-$N$ 
\begin{equation} 
\overline{G^\pm(E_1)G^\pm(E_2)} = \overline{G^\pm(E_1)} \cdot 
     \overline{G^\pm(E_2)} \;, 
\label{product}
\end{equation} 
we obtain 
\begin{equation} 
C(E_1,E_2) = \frac{1}{2\pi^2 N^2}\re\left[\overline{G^+(E_1)G^-(E_2)} 
     - \overline{G^+(E_1)} \cdot \overline{G^-(E_2)}\right] 
\label{C12} 
\end{equation} 
in the large-$N$ limit. Hence we shall focus our attention on computing the 
quantity 
\mbox{$\overline{G^+(E_1)G^-(E_2)}$}. 
Eq.~(\ref{product}) can be seen to follow from the triviality of the
large-$N$ saddle point (as discussed later on) when both imaginary parts
lie on the same side of the real axis. 

\renewcommand{\arraystretch}{0.5}
It is convenient to introduce the average energy 
\mbox{$E = \half(E_1+E_2)$} 
and the energy difference 
\linebreak 
\mbox{$\omega = E_1 - E_2$}. 
Then we can write 
\begin{equation} 
E_1 = E+i\tilde{\omega} \;, \quad 
E_2 = E-i\tilde{\omega} \;, 
\end{equation} 
where we define 
\mbox{$\tilde{\omega} = \half(\epsilon - i\omega)$}. 
Thus we have 
\begin{equation} 
\left( 
\begin{array}{cc} 
E^+_1 & 0     \\ 
0     & E^-_2 
\end{array} 
\right) = E{\opfone}_2 + i\tilde{\omega}L \;, \quad L = 
\left( 
\begin{array}{rr} 
+1 & 0  \\ 
0  & -1 
\end{array} 
\right) \;, 
\end{equation} 
and so 
\begin{equation} 
G^+(E_1)G^-(E_2) = \Tr\left(\frac{1}{E{\opfone}_2-H 
     +i\tilde{\omega}L}\right)_{11}
     {\cdot}\Tr\left(\frac{1}{E{\opfone}_2-H+i\tilde{\omega}L}\right)_{22} 
     \;. 
\label{G1G2} 
\end{equation} 
The symbol ${\opfone}_2$ is just the two-dimensional unit matrix.

\chapter{Superalgebra}
The supersymmetry formalism necessitates the use of anti-commuting (or 
Grassmann) numbers. Such numbers satisfy an anti-commutative product law, 
\mbox{$\eta_1\eta_2 = -\eta_2\eta_1$}. 
Consequently, they are nilpotent, 
\mbox{$\eta_1^2 = \eta_2^2 = 0$}. 
Complex conjugation can also be defined for Grassmann numbers. 
Our convention for complex conjugation is 
\begin{equation} 
(\eta_1\eta_2)^* = \eta_1^*\eta_2^* \;, \quad 
\eta^{**} = -\eta \;. 
\end{equation} 
We note that the product of two Grassmann numbers is a commuting number. 
Thus, a Grassmann algebra (or superspace) can be constructed by combining 
anti-commuting numbers with commuting (real or complex) ones. 
Linear algebra, analysis and topology can all be introduced on superspaces. 
For an exposition or review, one can consult 
Refs.~\citenum{Efetov}--\citenum{IZ}. 
Here, we shall restrict ourselves to defining the basic quantities and 
constructions needed for later use, and presenting our conventions. 
Because of their relevance to the path-integral formulation of 
elementary-particle theories, commuting degrees of freedom in a superspace are often 
called `bosonic', and anti-commuting ones `fermionic'. 
A differentiable manifold with both bosonic and fermionic coordinates is 
known as a supermanifold. The superspaces and supermanifolds that we 
shall encounter in the present application are ones with equal numbers of 
bosonic and fermionic degrees of freedom. This situation is sometimes 
described as a perfect $\mbox{Z}_2$-grading, and symmetries on such spaces 
are supersymmetries. 

Supermatrices have the form 
\begin{equation} 
A = \left( 
\begin{array}{cc} 
a     & \alpha \\ 
\beta & b 
\end{array}
\right) \;, 
\end{equation} 
where 
\mbox{$a,b$} 
are sub-matrices of commuting elements and 
\mbox{$\alpha,\beta$} 
are sub-matrices of anti-commuting elements. 
Let us label the elements of the sub-matrices by the indices 
\mbox{$p,p' = 1,2,\ldots$}, 
and the four blocks by indices 
\mbox{$\alpha,\alpha' = 0,1$} 
such that 
\begin{equation} 
A^{00}_{pp'} = a_{pp'} \;, \quad 
A^{01}_{pp'} = \alpha_{pp'} \;, \quad 
A^{10}_{pp'} = \beta_{pp'} \;, \quad 
A^{11}_{pp'} = b_{pp'} \;. 
\end{equation} 
In this scheme, where elements of $A$ are represented as 
\mbox{$A^{\alpha\alpha'}_{pp'}$}, 
the indices 
\mbox{$\alpha,\alpha'$} 
govern the grading of the matrix elements, such that 
\mbox{$\alpha+\alpha'$} 
even corresponds to commuting elements while 
\mbox{$\alpha+\alpha'$} 
odd to anti-commuting ones. We shall refer to 
\mbox{$\alpha,\alpha'$} 
as the `graded indices'. The interchange of supermatrix elements can now be 
compactly expressed as 
\begin{equation} 
A^{\alpha\alpha'}_{pp'}B^{\beta\beta'}_{qq'} = 
     (-1)^{(\alpha+\alpha')(\beta+\beta')} 
     B^{\beta\beta'}_{qq'}A^{\alpha\alpha'}_{pp'} \;. 
\end{equation} 
One commonly calls $A^{00}$ and $A^{11}$ the `boson-boson' and 
`fermion-fermion' blocks, respectively, while $A^{01}$ and $A^{10}$ are 
known as the `boson-fermion' and `fermion-boson' blocks, respectively. 

The graded trace (or supertrace) of a supermatrix $A$ is defined to be 
\begin{equation} 
\trg A = \tr a - \tr b \;, 
\end{equation} 
or, equivalently,
\begin{equation} 
\trg A = \sum_{p,\alpha} (-1)^\alpha A^{\alpha\alpha}_{pp} \;. 
\end{equation} 
The graded determinant (or superdeterminant) can be defined from the graded 
trace according to 
\begin{equation} 
\detg A = e^{\trg\ln A} \;. 
\end{equation} 
From this, one can show that 
\begin{equation} 
\detg A = \frac{\det (a - \alpha b^{-1} \beta)}{\det b} \;. 
\end{equation} 
Constructed in this way, the graded determinant has all the usual 
properties, such as 
\linebreak 
\mbox{$\detg A^{-1} = (\detg A)^{-1}$}, 
\mbox{$\detg (AB) = \detg A {\cdot} \detg B$}, 
etc. Also, for a supermatrix diagonal in the graded indices, 
\mbox{$A^{\alpha\alpha'}_{pp'} = \delta^{\alpha\alpha'} 
     A^{(\alpha)}_{pp'}$}, 
we have
\begin{equation}
\detg A = \frac{\det A^{(0)}}{\det A^{(1)}} \;. 
\end{equation} 
Consequently, 
\mbox{$\detg {\opfone} = 1$}, 
but does not exist for the zero supermatrix. 

Supermatrices act on supervectors $\varphi$, whose components we denote by 
\mbox{$\varphi^\alpha_p$}, 
so that 
\mbox{$\alpha = 0,1$} 
corresponds to commuting and anti-commuting elements, respectively. 
Thus, if $\varphi$ and $\psi$ are supervectors, then 
\begin{equation} 
\varphi^\alpha_p\psi^{\alpha'}_{p'} = (-1)^{\alpha\alpha'} 
     \psi^{\alpha'}_{p'}\varphi^\alpha_p \;. 
\end{equation} 

The transpose (sometimes referred to as supertranspose) of a supermatrix 
$A$ is defined by the requirement that 
\mbox{$\psi^{\rm T}A^{\rm T}\varphi = (A\psi)^{\rm T}\varphi$} 
for any supervectors 
\mbox{$\varphi,\psi$}. 
It follows that 
\begin{equation} 
A^{\rm T} = \left( 
\begin{array}{cc} 
a     & \alpha \\ 
\beta & b 
\end{array}
\right)^{\rm T} = 
\left( 
\begin{array}{rr} 
a^{\rm T}     & \beta^{\rm T} \\ 
-\alpha^{\rm T} & b^{\rm T} 
\end{array}
\right) \;.
\end{equation} 
This property can be equivalently expressed as 
\begin{equation} 
\left(A^{\rm T}\right)^{\alpha\alpha'}_{pp'} = (-1)^{\alpha(1+\alpha')} 
     A^{\alpha'\alpha}_{p'p} \;. 
\end{equation} 
We see that double transpose is not equivalent to the identity map. Instead, 
we have 
\begin{equation} 
\begin{array}{ccccccc}
A^{**} & = & kAk \;, & \quad\quad & \varphi^{**} & = & k\varphi \;, \\
A^{\rm TT} & = & kAk \;, & \quad\quad & A^{\dagger\dagger} & = & A \;,  
\end{array} 
\end{equation} 
where 
\mbox{$k^{\alpha\alpha'}_{pp'} = (-1)^\alpha\delta^{\alpha\alpha'} 
     \delta_{pp'}$} 
and the Hermitian conjugate is defined as the composition of complex 
conjugation and supertranspose, viz., 
\mbox{$A^\dagger = A^{\rm *T}$}.

\chapter{Supersymmetry Formalism} 
\vspace{-2ex} 
\section{Basic Ideas} 
We look for a generating function for traces of resolvents 
\mbox{$(E^\pm-H)^{-1}$}, 
where powers and products are obtained by differentiation with respect to a 
source. As a simple example, let us consider 
\begin{equation} 
Z(\varepsilon) = \frac{\det (E-H+\varepsilon{\opfone})}{\det (E-H)} \;. 
\label{Z} 
\end{equation} 
Then 
\begin{equation} 
\frac{dZ(\varepsilon)}{d\varepsilon} = \Tr\frac{1}{E-H+\varepsilon}\cdot 
     Z(\varepsilon) \;. 
\end{equation} 
All powers and products can be generated if one promotes 
\mbox{$E-H+\varepsilon{\opfone}$}, 
\mbox{$E-H$} 
to block-diagonal matrices by taking 
\mbox{$\varepsilon = \diag(\varepsilon_1,\varepsilon_2,\ldots,\varepsilon_M)$}, 
\mbox{$E = \diag(E_1,E_2,\ldots,E_M)$}, 
and differentiating with respect to the source components 
\mbox{$\varepsilon_k$}, 
\mbox{$k = 1,2,\ldots,M$}. 
The ratio of determinants appearing in Eq.~(\ref{Z}) ensures that the 
generating function is properly normalized, 
\mbox{$Z(0)=1$}. 
It follows immediately that if we can ensemble-average the generating 
function, then differentiation with respect to the source will 
automatically generate ensemble-averaged resolvent products, such as those 
required by Eq.~(\ref{G1G2}). 

For a Gaussian-distributed random Hamiltonian $H$, this ensemble average is 
easy to perform if its appearance in 
\mbox{$Z(\varepsilon)$} 
is purely exponential. 
Indeed, the central element of the present approach is the fact that if 
$H$ is an $N\times N$ random matrix belonging to any Gaussian distribution 
with zero mean, then 
\begin{equation} 
\overline{e^{-i\Tr HA}} = e^{-\frac{1}{2}\overline{(\Tr HA)^2}} 
\label{exp} 
\end{equation} 
for any fixed $N\times N$ matrix $A$. 
This result can be easily demonstrated as follows: Since 
\mbox{$\overline{H}=0$}, expanding the exponent yields 
\begin{eqnarray} 
\overline{e^{-i\Tr HA}} & = & \sum_{n=0}^\infty \frac{(-1)^n}{(2n)!} 
     \overline{(\Tr HA)^{2n}} 
\nonumber \\ 
& = & \sum_{n=0}^\infty \frac{(-1)^n}{(2n)!}{\cal N}_{\rm pair}(n) 
     \left[\overline{(\Tr HA)^2}\right]^n \;, 
\end{eqnarray} 
where we have appealed to Wick's theorem, which expresses the fact that for 
{\em any} Gaussian distribution, all moments factorize into second moments, 
i.e., the fully connected parts of all higher moments vanish. 
Thus we see that 
\mbox{${\cal N}_{\rm pair}(n)$} 
denotes the number of ways to divide $2n$ distinct objects into $n$ pairs, 
\begin{equation} 
{\cal N}_{\rm pair}(n) \quad = \quad (2n-1)!! 
     \quad = \quad \frac{(2n)!}{2^nn!} \;. 
\end{equation} 
Eq.~(\ref{exp}) follows immediately.
Such exponential dependence on $H$ 
can be achieved by expressing the determinants 
in the denominator and numerator of Eq.~(\ref{Z}) for the generating 
function as Gaussian integrals over commuting and anti-commuting variables, 
respectively. We shall now proceed to discuss this construction. 

Let $A_0$ be a non-singular $N\times N$ matrix whose Hermitian part is 
positive definite. Then one can express the inverse determinant of $A_0$ 
as a Gaussian integral in the holomorphic representation \cite{IZ}, given 
by 
\begin{equation} 
\frac{1}{\det A_0} = \int\prod_{k=1}^N\frac{dz_kdz_k^*}{2\pi i}\; 
     e^{-z^\dagger A_0z} \;, 
\label{bosonic}
\end{equation} 
where $z$ denotes a complex N-dimensional vector 
\mbox{$z = (z_1,z_2,\ldots,z_N)^{\rm T}$} 
and $z^\dagger$ is its Hermitian conjugate. 
There exists an analogous Gaussian-integral form for the determinant of 
any $N\times N$ matrix $A_1$ in terms of Grassmann variables \cite{IZ}. It 
is given by 
\begin{equation} 
\det A_1 = \int\prod_{\ell=1}^N (2\pi)\,d\eta_\ell^* d\eta_\ell \; 
     e^{-\eta^\dagger A_1\eta} \;, 
\label{fermionic}
\end{equation} 
assuming that the Grassmann integrals are normalized according to 
\begin{equation} 
\int d\eta_\ell\, \eta_\ell = \int d\eta_\ell^*\, \eta_\ell^* = 
     (2\pi)^{-1/2} \;. 
\label{45}
\end{equation} 
(We shall find this choice useful later on.) 
As a consistency condition, Grassmann integrals necessarily satisfy 
\begin{equation}
\int d\eta_\ell = \int d\eta_\ell^* = 0 \;. 
\label{zero}
\end{equation} 
The simplest way of seeing why these identities must hold is to note that
the value of an integral (being a measure) must be a c-number (i.e.\ an
even element of the Grassmann algebra); however, the expressions in
Eq.~(\ref{zero}) are of first order in the anti-commuting differentials 
\mbox{$d\eta_\ell, d\eta_\ell^*$} 
(i.e.\ odd elements of the Grassmann algebra). 
Thus, zero remains as the only consistent value.
In the exponent above, $\eta$ denotes an $N$-dimensional vector of 
Grassmann components 
\mbox{$\eta = (\eta_1,\eta_1,\ldots,\eta_N)^{\rm T}$}. 
We remark that the multi-dimensional Gaussian integrals (\ref{bosonic}) and 
(\ref{fermionic}) form
the cornerstones of the path-integral formulations for systems of bosons
and fermions, respectively. Derivations of these identities appear in most
modern textbooks on quantum field theory and many-body physics, such as
Ref.~\citenum{IZ}. Nevertheless, for the sake of completeness, we present
derivations of Eqs.~(\ref{bosonic}) and (\ref{fermionic}) in Appendix~A.

We can combine the matrices $A_0, A_1$ into a block-diagonal matrix 
\mbox{$A = \diag(A_0,A_1)$}. 
This matrix can formally be considered to be a supermatrix (with vanishing 
anti-commuting blocks). Then 
\begin{equation} 
\detg A = \frac{\det A_0}{\det A_1} \;, 
\end{equation}
and we can write 
\begin{equation} 
\detg^{-1}A = \int\prod_{k=1}^N i\,dz_kdz_k^* 
     \prod_{\ell=1}^N d\eta_\ell d\eta_\ell^* \; 
     e^{-\varphi^\dagger A\varphi} \;, 
\label{sgauss}
\end{equation}
where $\varphi$ is the $2N$-dimensional supervector 
\mbox{$\varphi = (z,\eta)^{\rm T}$}. 
It is straightforward to show that the representation (\ref{sgauss}) 
generalizes to supermatrices with non-vanishing anti-commuting blocks. 

\section{Generating Function} 
Let us consider a generating function given by 
\begin{eqnarray}
Z(\varepsilon) & = & \Detg^{-1}[D+J(\varepsilon)] 
\nonumber \\
& = & \exp\left\{-\Trg\ln [D+J(\varepsilon)]\right\} \;,
\label{genfn}
\end{eqnarray}
where the symbols `Detg' and `Trg' denote the graded determinant and graded
trace as defined in the supersymmetric formalism of Ref.~\citenum{VWZ} 
and the foregoing discussion. 
Here, the inverse propagator $D$ has been extended to a 
$4N\times4N$ supermatrix, 
\begin{equation}
D = (E-H){\opfone}_4 +i\tilde{\omega} L \;, 
\label{dsuper}
\end{equation}
where 
\mbox{$L_{pp'}^{\alpha\alpha'} = (-1)^{p+1}\delta_{pp'}
     \delta^{\alpha\alpha'}$} 
is the diagonal supermatrix that distinguishes between advanced and retarded
parts of $D$. 
We have 
\mbox{$D_{pp'}^{\alpha\alpha'} = \delta^{\alpha\alpha'} 
     [\diag (D_+^{\vphantom{\dagger}},D_+^\dagger)]_{pp'}$},
where 
\begin{equation} 
D_+ = E-H+i\tilde{\omega} \;, \quad D_- = E-H-i\tilde{\omega} 
\end{equation} 
so that 
\mbox{$D_- = D_+^\dagger$}, 
and 
\mbox{$p,p'=1,2$}, 
with
\mbox{$p=1$} referring to the retarded block and 
\mbox{$p=2$} to the advanced block.
The indices 
\mbox{$\alpha,\alpha' = 0,1$} 
determine the grading (with 
\mbox{$\alpha=0$} 
for the commuting (bosonic) components and 
\mbox{$\alpha=1$}
for the anti-commuting (fermionic) components). 
Our notational conventions in writing traces and determinants are as follows:
Those beginning with lower-case letters (e.g.\ trg, detg) imply 
summation only over graded indices 
\mbox{$\alpha,\alpha'$} 
and/or block indices 
\mbox{$p,p'$}, 
while those beginning with an upper-case letter (e.g.\ Trg, Tr, Detg) 
indicate that there is also (or perhaps solely) a summation over the level 
indices 
\mbox{$\mu,\nu$}. 

The source supermatrix $J(\varepsilon)$ depends on a set of parameters 
$\varepsilon_m$ labelled by some (multi)-index $m$, and is taken to have the
general form 
\begin{equation}
J(\varepsilon) = \varepsilon_m M_m 
\label{412}
\end{equation}
for some set of $4N\times4N$ supermatrices $M_m(\mu,\nu)$. 
It then follows that 
\begin{equation}
\left. \frac{\partial^2}{\partial\varepsilon_m\partial\varepsilon_n} 
     Z(\varepsilon)\right|_{\varepsilon=0} = 
     \Trg M_mD^{-1}{\cdot}\Trg M_nD^{-1} + 
     \Trg M_mD^{-1}M_nD^{-1} \;. 
\label{diffj} 
\end{equation}
For the problem at hand, we shall make the choice of source matrix
(here independent of the level indices 
\mbox{$\mu,\nu$}) 
\begin{equation} 
J^{\alpha\alpha'}_{pp'}(\varepsilon_1,\varepsilon_2) = 
     \sum_{m=1,2}\varepsilon_m I_{pp'}(m) k^{\alpha\alpha'} \;, 
\end{equation} 
where
\begin{equation}
I_{pp'}(m) = \delta_{pp'}\delta_{pm} 
\label{proj}
\end{equation}
is a projector onto the $p$-block with 
\mbox{$p = m$} 
and 
\mbox{$k^{\alpha\alpha'} = (-1)^\alpha\delta^{\alpha\alpha'}$}. 
In this case, we have 
\mbox{$J = \varepsilon_m M_m$} 
with the identification 
\begin{equation} 
M_m = I(m)\otimes k \;, \quad m \in \{1,2\} \;. 
\label{Mm} 
\end{equation} 
With $M_m$ given by Eq.~(\ref{Mm}), 
the second term on the RHS of Eq.~(\ref{diffj}) vanishes if we
differentiate with respect to 
\mbox{$\varepsilon_m = \varepsilon_1$} 
and 
\mbox{$\varepsilon_n = \varepsilon_2$},  
and we see that 
\begin{equation} 
\Trg M_mD^{-1} = \left\{ 
\begin{array}{ccl} 
2\Tr(D_+)^{-1}  & \quad\mbox{for}\quad & m=1 \\ 
2\Tr(D_-)^{-1}  & \quad\mbox{for}\quad & m=2 \;, 
\end{array} 
\right. 
\end{equation} 
noting that 
\mbox{$\trg k = 2$} 
here. Therefore, 
\begin{equation} 
G(E_1^+)G(E_2^-) = \left. \frac{1}{4} 
     \frac{\partial^2}{\partial\varepsilon_1\partial\varepsilon_2} 
     Z(\varepsilon)\right|_{\varepsilon=0} 
\end{equation} 
We shall sometimes employ the obvious notation 
\mbox{$G(E^\pm) \equiv G^\pm(E)$}.  

\section{Ensemble Average} 
As discussed in the previous section, the generating function can be 
expressed as a Gaussian superintegral, 
\begin{equation}
Z(\varepsilon) = \int{\cal D}\varphi{\cal D}\overline{\varphi}\, 
     e^{i{\cal L}_1(\varphi;J)} \;, 
\label{gsuper} 
\end{equation} 
where 
\begin{equation} 
{\cal L}_1(\varphi;J) = 
\sum_{p,\alpha}\langle\overline{\varphi}_p^\alpha, 
     [(D+J)\varphi]_p^\alpha\rangle \;,
\label{L1}
\end{equation}
and the measure 
\mbox{${\cal D}\varphi {\cal D}\overline{\varphi}$} 
denotes 
\begin{equation} 
{\cal D}\varphi {\cal D}\overline{\varphi} = \prod_{\mu,p,\alpha} 
     i^\alpha d\varphi^\alpha_p(\mu) 
     d\overline{\varphi}^\alpha_p(\mu) \;. 
\end{equation} 
We employ the notation
\begin{equation}
\langle F,G \rangle \equiv \sum_\mu F(\mu)G(\mu) \;,
\end{equation} 
and 
\mbox{$\varphi_p^\alpha(\mu)$} 
is a four-component supervector field. 
The adjoint supervector is defined by 
\mbox{$\overline{\varphi} = \varphi^\dagger s$}
with
\begin{equation}
s_{pp'}^{\alpha\alpha'} = s_p^\alpha\delta^{\alpha\alpha'}\delta_{pp'}\;, 
     \quad\quad s^\alpha_p = (-1)^{(1-\alpha)(1+p)} \;. 
\label{sdef}
\end{equation}
The presence of the supermatrix $s$ in the definition of the adjoint ensures
the convergence not only of Eq.~(\ref{gsuper}) but also 
of the final integral representation of the ensemble-averaged
generating function as a supermatrix non-linear $\sigma$-model and ensures the
correct combination of compact and non-compact symmetries therein \cite{VZ}. 
By construction, we have 
\mbox{$Z(0)=1$}. 

The ensemble-averaged generating function reads 
\begin{equation}
\overline{Z(\varepsilon)} = \int{\cal D}\varphi{\cal D}\overline{\varphi}\, 
     \overline{\exp\Bigl\{-i\sum_{p,\alpha}\langle\overline{\varphi}_p^\alpha, 
     (H\varphi)_p^\alpha\rangle\Bigr\}}^H
     \exp\Bigl\{i\sum_{p,\alpha}\langle\overline{\varphi}_p^\alpha, 
     [(E{\opfone}+i\tilde{\omega}L+J)\varphi]_p^\alpha\rangle\Bigr\} \;,
\label{AVZ}
\end{equation}
and we can write 
\begin{equation} 
\overline{\exp\Bigl\{-i\sum_{p,\alpha}\langle\overline{\varphi}_p^\alpha, 
     (H\varphi)_p^\alpha\rangle\Bigr\}}^H = 
     \overline{\exp\Bigl\{-i\Tr_\mu H{\cal S}\Bigr\}}^H \;, 
\end{equation} 
where we have introduced the ordinary $N\times N$ matrix in the level 
indices 
\mbox{$\mu,\nu$}, 
\begin{equation} 
{\cal S}(\nu,\mu) = \sum_{p,\alpha} 
     \overline{\varphi}^\alpha_{p}(\mu)\varphi^\alpha_{p}(\nu) \;. 
\end{equation} 
We can use Eq.~(\ref{exp}) to obtain the result
\begin{eqnarray} 
\overline{\exp\Bigl\{-i\Tr_\mu H{\cal S}\Bigr\}}^H  & = &  
     \exp\Bigl\{-\half\overline{(\Tr_\mu H{\cal S})^2}^H\Bigr\} 
\nonumber \\ 
& = & \exp\Bigl\{-\frac{\lambda^2}{2N}\Tr_\mu {\cal S}^2\Bigr\}
\nonumber \\ 
& = & \exp\Bigl\{-\frac{\lambda^2}{2N}\trg S^2\Bigr\} \;, 
\end{eqnarray} 
where 
\begin{equation}
S^{\alpha\alpha'}_{pp'} \equiv \sum_\mu\varphi^\alpha_{p}(\mu) 
     \overline{\varphi}^{\alpha'}_{p'}(\mu) 
\end{equation}
is a supermatrix, and it is straightforward to show that 
\begin{equation} 
\Tr_\mu {\cal S}^2 = \sum_{\mu,\nu} {\cal S}(\mu,\nu) {\cal S}(\nu,\mu) 
     = \trg S^2 \;. 
\end{equation} 
The ensemble-averaged generating function can now be expressed as 
\begin{equation}
\overline{Z(\varepsilon)} = \int{\cal D}\varphi{\cal D}\overline{\varphi}\, 
     e^{i{\cal L}_2(S;J)} \;, 
\label{ZL2} 
\end{equation} 
where 
\begin{equation} 
i{\cal L}_2(S;J) = -\frac{\lambda^2}{2N}\trg S^2 + 
     i\trg (E{\opfone}+i\tilde{\omega}L+J)S \;. 
\end{equation} 

\section{Hubbard-Stratonovich Transformation} 
Ensemble averaging of the generating function
\mbox{$Z(\varepsilon)$} 
has introduced a term quartic in the supervector $\varphi$ into the 
exponent. Consequently, the supervector integration can no longer be 
performed exactly. However, considerable simplification can be achieved 
through a Hubbard-Stratonovich transformation which serves to eliminate the 
quartic interaction in favour of new (composite) degrees of freedom 
comprised of $4\times 4$ supermatrices $\sigma$, which couple to the 
dyadic form 
\mbox{$\sum_{\mu}\varphi(\mu)\overline{\varphi}(\mu)$}. 
It turns out that the expectation of (an appropriate graded trace over) 
$\sigma$ is proportional to the average 
density of states, and hence non-zero. We make this explicit in Appendix~C. 
Thus, it is analogous to an order-parameter field signalling the occurrence of 
spontaneous symmetry breaking, the density of states being the actual order 
parameter. Moreover, 
the $\sigma$ supermatrices can easily be decomposed into massless (Goldstone)
modes and massive modes which conveniently decouple in the large-$N$ limit, 
and leave us with a theory of interacting Goldstone modes. 
These are the degrees of freedom in terms of which all quantities in the final 
effective theory can be expressed, and the Hubbard-Stratonovich 
transformation serves to extract them from the original ones  
\mbox{$\varphi(\mu)$}.  
Hence, another advantage of this procedure is that the number of degrees of 
freedom is reduced by a factor $N$, since the $\sigma$ modes are 
independent of the level indices 
\mbox{$\mu,\nu$}.   

To implement the Hubbard-Stratonovich transformation, we follow 
Ref.~\citenum{PS} in defining 
\begin{equation} 
i{\cal L}^\#_2(S) = -\frac{\lambda^2}{2N}\trg S^2 - 
     \tilde{\omega}_{\rm r}\trg LS  
\end{equation} 
and 
\begin{equation} 
W(\sigma,S) = -{\cal L}_2^\#\left(S-\frac{iN}{\lambda}\sigma\right) \;.
\label{WS}
\end{equation} 
where 
\mbox{$\tilde{\omega}_{\rm r} \equiv \re \tilde{\omega}$}, 
and correspondingly 
\mbox{$\tilde{\omega}_{\rm i} \equiv \im \tilde{\omega}$}.
Then we can write 
\begin{equation}
\overline{Z(\varepsilon)} = \int{\cal D}\sigma \, 
     e^{i{\cal L}_3(\sigma;J)} \;, 
\label{ZL3} 
\end{equation} 
where 
\begin{equation}
e^{i{\cal L}_3(\sigma;J)} = \int{\cal D}\varphi{\cal D}\overline{\varphi}\, 
     e^{iW(\sigma,S)+i{\cal L}_2(S;J)} \;, 
\label{L3} 
\end{equation} 
assuming that 
\begin{equation} 
\int {\cal D}\sigma\, e^{iW(\sigma,S)} = 
     \int {\cal D}\sigma\, e^{iW(\sigma,0)} = 1 \;, 
\label{Wint} 
\end{equation} 
i.e.\ shifts are allowed. 
The choice of 
\mbox{$W(\sigma,S)$} 
given in Eq.~(\ref{WS}) serves to eliminate the quartic dependence on 
$\varphi$ in the exponent of the generating function implied in 
\mbox{${\cal L}_2(S;J)$}, since 
\begin{equation} 
iW(\sigma,S) + i{\cal L}^\#_2(S) = -\frac{N}{2}\trg\sigma^2 
     - \frac{iN\tilde{\omega}_{\rm r}}{\lambda}\trg\sigma L 
     - i\lambda\trg\sigma S \;. 
\end{equation} 
The remaining quadratic dependence on $\varphi$ is amenable to exact 
integration, which yields 
\begin{eqnarray} 
e^{i{\cal L}_3(\sigma,J)} & = & \exp\Bigl\{-\frac{N}{2}\trg\sigma^2 
     - \frac{iN\tilde{\omega}_{\rm r}}{\lambda}\trg\sigma L\Bigr\}
     \int{\cal D}\varphi{\cal D}\overline{\varphi}\, 
     \exp\Bigl\{i\sum_{p,\alpha}\langle\overline{\varphi}_p^\alpha, 
     [(E{\opfone}-\tilde{\omega}_{\rm i}L+J-\lambda\sigma)
     \varphi]_p^\alpha\rangle\Bigr\} \;.
\nonumber \\ 
& = & \exp\Bigl\{-\frac{N}{2}\trg\sigma^2 
     - \frac{iN\tilde{\omega}_{\rm r}}{\lambda}\trg\sigma L
     - N\trg\ln\left[E{\opfone} - \tilde{\omega}_{\rm i}L + J 
     -\lambda\sigma\right]\Bigr\} \;.
\end{eqnarray} 
At this stage, it is advantageous to perform the shift of integration 
matrix 
\mbox{$\sigma\mapsto\sigma'$} 
defined by 
\begin{equation} 
-\lambda\sigma' = -\tilde{\omega}_{\rm i}L + J - \lambda\sigma 
\end{equation} 
in order to move the source matrix and $\tilde{\omega}_{\rm i}$ 
out of the logarithm. 
Noting that 
\begin{eqnarray} 
\trg\sigma^2 & = & \trg{\sigma'}^2 + \frac{1}{\lambda^2}\left[ 
     -2\lambda\tilde{\omega}_{\rm i}\trg\sigma' L 
     + 2\lambda\trg\sigma' J 
     - 2\tilde{\omega}_{\rm i}\trg LJ 
     + \trg J^2 \right] \;, 
\nonumber \\ 
\trg\sigma L & = & \trg\sigma' L + \frac{1}{\lambda} \trg LJ \;, 
\end{eqnarray} 
we are left with 
\begin{equation}
\overline{Z(\varepsilon)} = \int{\cal D}\sigma \, 
     \exp\Bigl\{-\frac{N}{2}\trg\sigma^2 
     - \frac{iN\tilde{\omega}}{\lambda}\trg\sigma L
     - \frac{N}{\lambda}\trg\left(\sigma+i\lambda^{-1}\tilde{\omega}L\right)J
     - \frac{N}{2\lambda^2}\trg J^2 
     - N\trg\ln\bigl[E{\opfone} -\lambda\sigma\bigr]\Bigr\} \;, 
\label{zsigma}
\end{equation} 
where we have dropped the primes on the $\sigma'$, and we observe that 
\mbox{$\tilde{\omega} = \tilde{\omega}_{\rm r} + i\tilde{\omega}_{\rm i}$}
has been reconstituted. 

\section{Integration Supermanifold} 
In the absence of symmetry breaking 
\mbox{$(\tilde{\omega} = J = 0)$},
\mbox{${\cal L}_1(\varphi;J)$} 
is invariant under linear transformations 
\mbox{$\varphi \mapsto T\varphi$} 
which preserve the bilinear form 
\mbox{$\overline{\varphi}\varphi$}. 
These $4\times 4$ supermatrices must satisfy 
\begin{equation} 
T^{-1} = sT^\dagger s \;, 
\label{sTs} 
\end{equation} 
and constitute a supergroup with compact ---  SU(2) --- and non-compact 
--- SU(1,1) --- 
bosonic subgroups. 
The transformation property induced on $S$ is given by 
\begin{equation} 
S \mapsto TST^{-1} \;. 
\label{TST} 
\end{equation}  
It is also useful to note that 
\mbox{$S^\dagger = sSs$}. 
The generating function expressed in terms of the supermatrices $\sigma$, 
as given by Eq.~(\ref{ZL3}), should also possess invariance under these 
transformations in the absence of symmetry breaking. The structure of 
\mbox{$W(\sigma,S)$} 
then implies that $\sigma$ should also transform like $S$, i.e.\ 
\mbox{$\sigma \mapsto T\sigma T^{-1}$}. 
Hence, the domain of integration over the supermatrices $\sigma$ should 
span a space invariant under 
\mbox{$\sigma \mapsto T\sigma T^{-1}$} \cite{PS,SW}. 
Such a space is clearly furnished by matrices of the form 
\mbox{$\sigma = T^{-1}P_{\rm d}T$}, 
where the $P_{\rm d}$ 
are diagonal matrices with real boson-boson  and imaginary fermion-fermion 
elements. 
% Such matrices also satisfy 
% \mbox{$\sigma^\dagger = s\sigma s$}. 
However, this choice still does not lead to a convergent integral (\ref{Wint}) 
over $\sigma$. 
To remedy this, we can add an imaginary part to $P_{\rm d}$ according to 
\begin{equation} 
\sigma = T^{-1}(P_{\rm d} + r{\opfone}- i\Delta L)T \;, 
\label{TPT} 
\end{equation} 
where $\Delta$ is an arbitrary positive constant. It is advantageous to 
choose $\Delta$ such that a saddle point of 
\mbox{${\cal L}_3(\sigma;0)$} 
lies within the $\sigma$-integration manifold. This criterion will fix 
$\Delta$ uniquely. 
We have also included a real term proportional to the unit matrix because, 
as we shall see, though not required for convergence, its presence is also 
necessary to allow the saddle point to lie within the $\sigma$-integration 
manifold. 

\renewcommand{\arraystretch}{0.5} 
It is convenient to re-express Eq.~(\ref{TPT}) in a slightly different form.
We perform a coset decomposition of the supergroup elements 
\mbox{$T = RT_0$} 
where the $R$ span the subgroup that commutes with $L$. Then we have 
\begin{equation} 
\sigma = T_0^{-1}(\delta P + r{\opfone} - i\Delta L)T_0 \;, 
\label{TPT0} 
\end{equation} 
where 
\mbox{$\delta P = R^{-1}P_{\rm d}R$}. 
The supermatrices $\delta P$ are block-diagonal in the indices 
\mbox{$(p,p')$}, 
and have real boson-boson and imaginary fermion-fermion eigenvalues. 
% and satisfy \mbox{$P^\dagger = sPs$}. 
Now, the generators of the coset elements $T_0$ have vanishing entries in 
both of the diagonal $p$-blocks. Thus, we can write 
\begin{equation} 
T_0 = \exp\left( 
\begin{array}{cc} 
0             & iw \\ 
i\overline{w} & 0 
\end{array} 
\right) \;,
\label{Tww}
\end{equation} 
where, from Eq.~(\ref{sTs}), we deduce the relationship 
\mbox{$\overline{w} = -kw^\dagger$} 
between these two $2\times 2$ supermatrices. 
It follows immediately that 
\mbox{$LT_0L = T_0^{-1}$}. 

If we write 
\mbox{$\sigma = T_0^{-1}PT_0$}, 
then the supermatrices $P$ represent massive modes which decouple in the 
large-$N$ limit. To achieve this decoupling, we shall observe that one can 
choose 
\mbox{$P = \sigma_0 + \delta P$}, 
where $\sigma_0$ is the unique diagonal saddle point that lies within the
integration manifold, in which case $\delta P$ are interpreted as 
the massive fluctuations around it. 
Thus we have 
\mbox{$\sigma = \sigma_{\rm G} + T^{-1}\delta PT$} 
with 
\mbox{$\sigma_{\rm G} = T^{-1}\sigma_0T$}. 
The massive fluctuations can be integrated out in the limit 
\mbox{$N\to\infty$}, 
which leads to an expression for 
\mbox{$\overline{Z(\varepsilon)}$} 
identical with Eq.~(\ref{zsigma}) except for the replacement 
\mbox{$\sigma \to \sigma_{\rm G}$} 
everywhere. 

\section{Convergence} 
Let us now consider the convergence properties of the $\sigma$-integral in 
Eq.(\ref{Wint}), setting 
\mbox{$S=0$}. 
We have 
\begin{eqnarray} 
iW(\sigma,0) & = & -\frac{N}{2}\trg\Bigl(\sigma + 
     \frac{i\tilde{\omega}_{\rm r}}{\lambda}L\Bigr)^2 
\nonumber \\ 
& = & -\frac{N}{2}\Bigl[\trg (P_{\rm d} + r{\opfone} - i\Delta L)^2  + 
     \frac{2i\tilde{\omega}_{\rm r}}{\lambda}\trg P_{\rm d}TLT^{-1} 
     + \frac{\Delta\tilde{\omega}_{\rm r}}{\lambda} \trg LTLT^{-1} 
     \Bigr] \;. 
\label{W0}
\end{eqnarray} 
Integration over the elements of $P_{\rm d}$ is clearly convergent: The 
boson-boson elements are real, while the fact that the fermion-fermion 
elements are imaginary compensates for the minus sign coming from the 
graded trace. Thus 
\mbox{$\trg P_{\rm d}^2$} 
is positive definite, and the shift by the constant term 
\mbox{$r{\opfone}-i\Delta L$} 
does not affect convergence properties. 
Integration over the anti-commuting blocks of $T$ cannot lead to any 
divergence, and neither can integration over the fermion-fermion block of 
$T$ since it spans the group SU(2), the parameters of which are all bounded 
(being real angles). On the other hand, the boson-boson block of $T$ spans 
the non-compact group SU(1,1), which has one unbounded real parameter. 
However, the final term in Eq.~(\ref{W0}) furnishes the damping which 
ensures convergence, provided 
\mbox{$\Delta\tilde{\omega}_{\rm r}/\lambda$} 
is positive. 
The middle term in Eq.~(\ref{W0}) does not cause any problems. 
In the boson-boson block, it is essentially imaginary and so gives an 
oscillating contribution to the integral. 
On the other hand, the imaginary elements of the fermion-fermion block 
combine with the imaginary prefactor to give rise to a potentially divergent 
contribution. However, since the fermion-fermion block of 
\mbox{$T^{-1}LT$}
is bounded, this term cannot compete with the damping due to the 
fermion-fermion part of 
\mbox{$\trg P_{\rm d}^2$}. 
To this end, we note that 
\begin{equation} 
\trg P_{\rm d}^2 + \frac{2i\tilde{\omega}_{\rm r}}{\lambda} 
     \trg P_{\rm d}TLT^{-1} = 
     \trg \Bigl[P_{\rm d} + \frac{i\tilde{\omega}_{\rm r}}{\lambda}
     TLT^{-1}\Bigr]^2 \;. 
\end{equation} 
This would not be so if the fermion-fermion block of $T$ were non-compact. 
Furthermore, a non-compact fermion-fermion block, which would result from 
taking 
\mbox{$s=L$} 
in Eq.~(\ref{sdef}), 
would also cause the third term of Eq.~(\ref{W0}) to give rise to a 
divergent integral, owing to the relative minus sign between the boson-boson
and fermion-fermion blocks in the graded trace. 
On the other hand, it is clear that the boson-boson block of $T$ cannot be 
arranged to be compact as well as this would necessitate choosing 
\mbox{$s={\opfone}$}, 
which in turn would lead to a manifestly divergent integral over 
$\varphi$ in Eq.~(\ref{gsuper}). 
Including non-zero $S$ does not alter the arguments presented above. 
The general conclusions which can be drawn are that the non-zero boson-boson 
elements of $P_{\rm d}$ must be real and the fermion-fermion elements 
imaginary, while the boson-boson block of $T$ must be non-compact and the 
fermion-fermion block compact. Convergence cannot be achieved in any other 
way. 

\section{Saddle-Point Equation} 
To determine the saddle points of the exponent of 
\mbox{$\overline{Z(\varepsilon)}$}, 
as it appears in Eq.~(\ref{zsigma}), we neglect $\tilde{\omega}$ (as it 
is assumed to be of order 
\mbox{$O(N^{-1})$})
and the source matrix $J$ (which can be considered as infinitesimal). 
The saddle-point equation then reads 
\begin{equation} 
\frac{\delta}{\delta\sigma}\left[\half\trg\sigma^2 + 
     \trg\ln (E\opfone - \lambda\sigma)\right] = 0 \;, 
\end{equation} 
or equivalently, 
\begin{equation} 
\lambda\sigma^2 - E\sigma + \lambda = 0 \;. 
\label{saddle}
\end{equation} 
The unique diagonal solution of this equation assuming a form consistent 
with Eq.~(\ref{TPT}), namely 
\linebreak 
\mbox{$\sigma_0 = r{\opfone} - i\Delta L$}, 
corresponds to 
\begin{equation} 
r = \frac{E}{2\lambda} \;, \quad 
\Delta = \sqrt{1-\left(\frac{E}{2\lambda}\right)^2} \;, 
\label{semi}
\end{equation} 
and we note the consistency condition 
\mbox{$|E| \le 2\lambda$}. 
As a function of $E$, $\Delta(E)$ reproduces Wigner's semicircle law; it 
is in fact proprtional to the average density of states. We can use the 
normalization condition (\ref{unity}) to deduce that 
\mbox{$\overline{\rho(E)} = \Delta(E)/(\pi\lambda)$}. 
It follows immediately that the mean level spacing at 
\mbox{$E=0$} 
is given by 
\mbox{$d = \pi\lambda/N$}. 

\section{Decoupling of Massive Modes} 
If we scale the 
\mbox{$\delta P$} 
such that 
\mbox{$\sigma = \sigma_{\rm G} + N^{-1/2}T^{-1}\delta P'T$}, 
then we have 
\begin{equation} 
\trg\sigma^2 = \trg\sigma_{\rm G}^2 + N^{-1}\trg(\delta P')^2 
     + 2N^{-1/2}\trg\sigma_0\delta P' \;, 
\end{equation} 
and 
\begin{equation} 
\trg\ln(E-\lambda\sigma) = \trg\ln(E-\lambda\sigma_{\rm G}) + 
     \trg\ln(1-N^{-1/2}\sigma_0\delta P') 
\end{equation} 
by the saddle-point equation 
\mbox{$\sigma_0 = \lambda(E-\lambda\sigma_0)^{-1}$}. 
Thus, 
\begin{eqnarray} 
-\frac{N}{2}\trg\sigma^2 - N\trg\ln(E-\lambda\sigma) & & \\
& & \hspace{-4.0cm} = \quad 
     -\frac{N}{2}\trg\sigma_{\rm G}^2 - N\trg\ln(E-\lambda\sigma_{\rm G}) 
     - \frac{1}{2}\trg(\delta P')^2 + \frac{1}{2}\trg\sigma_0\delta P' 
     \sigma_0\delta P' + O(N^{-1/2}) 
\nonumber \\ 
& & \hspace{-4.0cm} = \quad 
     -\frac{N}{2}\trg\sigma_{\rm G}^2 - N\trg\ln(E-\lambda\sigma_{\rm G}) 
     - \frac{1}{2}\trg\left[1+\half(\Delta^2-r^2)+ir\Delta L\right] 
     (\delta P')^2 + O(N^{-1/2}) \;. 
\nonumber 
\end{eqnarray} 
Also, 
\begin{eqnarray} 
\trg\sigma L & = & \trg\sigma_{\rm G}L + N^{-1/2}\trg\delta P' TLT^{-1} 
\nonumber \\ 
& {\displaystyle \asym{N\to\infty}} & \trg\sigma_{\rm G}L \;, 
\end{eqnarray} 
since, given that 
\mbox{$N\tilde{\omega} \sim O(N^0)$}, 
the latter term will produce only non-leading contributions. Similarly, 
\begin{eqnarray} 
\trg\sigma J & = & \trg\sigma_{\rm G}J + N^{-1/2}\trg\delta P' TJT^{-1} 
\nonumber \\ 
& {\displaystyle \asym{N\to\infty}} & \trg\sigma_{\rm G}J \;, 
\end{eqnarray} 
since source differentiations on the latter term will produce contributions 
suppressed by 
\mbox{$O(N^{-1/2})$} 
compared with those coming from the first term. 
Hence, we obtain 
\begin{equation} 
i{\cal L}_3(\sigma;J) \asym{N\to\infty} -\half\trg\left[ 
     1+\half(\Delta^2-r^2)+ir\Delta L\right](\delta P')^2
     + i{\cal L}_3(\sigma_{\rm G};J) \;. 
\end{equation} 
We see that the variables $\sigma_{\rm G}$ and $\delta P'$ decouple in the 
large-$N$ limit. 

We note that the space spanned by the supermatrices $\sigma$ is a linear 
space. Thus, the integration measure can be taken simply to be  
\begin{equation} 
{\cal D}\sigma = \prod_{\alpha,\alpha' \atop p,p'} 
     d\sigma_{pp'}^{\alpha\alpha'} \;, 
\end{equation} 
in which case, complex conjugates of matrix elements, such as 
\mbox{$(\sigma_{pp'}^{\alpha\alpha'})^*$}, 
should be expressed in terms of the original variables, viz.\ 
\mbox{$\sigma_{p'p}^{\alpha'\alpha}$}. 
Let us now recall that we have the decomposition 
\mbox{$\sigma = T^{-1}_0PT_0$}, 
where 
\mbox{$P = \delta P + r{\opfone} - i\Delta L$}. 
If we take the independent elements of $P$ and the coset 
supermatrices $T_0$ as our integration variables, then the measure on the 
$\sigma$ becomes \cite{VZ}
\begin{equation} 
{\cal D}\sigma = I(P){\cal D}P{\cal D}\mu(T_0) \;, 
\end{equation} 
where
\mbox{${\cal D}\mu(T_0)$} 
is the invariant measure on the $T_0$-coset manifold, 
\begin{equation} 
{\cal D}P = {\cal D}P_1{\cdot}{\cal D}P_2 \;, 
     \quad {\cal D}P_p = -i\prod_{\alpha,\alpha'} 
     dP_p^{\alpha\alpha'} \;, 
\end{equation} 
for 
\mbox{$p=1,2$}, 
and 
\begin{equation} 
I(P) = \biggl[\frac{(\Lambda_1^0-\Lambda_2^0) 
     (\Lambda_1^1-\Lambda_2^1)}{(\Lambda_1^0-\Lambda_2^1)
     (\Lambda_1^1-\Lambda_2^0)}\biggr]^2 \;, 
\end{equation} 
where $\Lambda_p^\alpha$, 
\mbox{$\alpha = 0,1$}, 
are the two eigenvalues of $P_p$, for 
\mbox{$p=1,2$}, 
respectively. Now $\sigma$, and hence $P$, are diagonalized by 
\begin{equation} 
\sigma = T^{-1}(P_{\rm d} + r{\opfone} - i\Delta L)T \;. 
\end{equation} 
Thus we see that 
\begin{equation} 
\diag(\Lambda_1^0,\Lambda_1^1,\Lambda_2^0,\Lambda_2^1) = 
     N^{-1/2}P_{\rm d}' + r{\opfone} -i\Delta L 
\end{equation} 
after scaling, i.e. 
\begin{equation} 
\Lambda_p^\alpha = r +i(-1)^p\Delta + O(N^{-1/2}) \;. 
\end{equation} 
Consequently, 
\mbox{$I(P) = 1 + O(N^{-1/2})$}. 
The massive fluctuations 
\mbox{$\delta P_1',\delta P_2'$} 
can now be easily integrated out in the large-$N$ limit, to yield 
\begin{equation} 
\overline{Z(\varepsilon)} \asym{N\to\infty} K_1K_2\int {\cal D}\mu(T_0)\,
     e^{i{\cal L}_3(\sigma_{\rm G};J)} \;, 
\end{equation} 
where 
\begin{equation} 
K_p = \int {\cal D}P_p\, e^{-\half c_p\trg(\delta P_p')^2} \;, 
\label{465}
\end{equation} 
with 
\mbox{$c_p = 1+\half(\Delta^2-r^2)-i(-1)^pr\Delta$} 
for 
\mbox{$p=1,2$}. 
Clearly, 
\mbox{$K_1 = K_2 = 1$}. 
We show this explicitly in Appendix~B. 
Therefore, we see that the large-$N$ limit is obtained by setting the 
massive fluctuations 
\mbox{$\delta P$} 
to zero and dropping the $P$-integration, which effects the substitution 
\mbox{$\sigma \to \sigma_{\rm G}$}, 
as mentioned previously. 
In other words, we treat the massive $P$-sector at tree level. 

We note that 
\mbox{$\trg\sigma_{\rm G}^2 = 0$}, and by the saddle-point equation 
(\ref{saddle}), 
\begin{equation} 
\trg\ln(E{\opfone} - \lambda\sigma_{\rm G}) = 
     -\trg\ln(\lambda^{-1}\sigma_{\rm G}) = 0 \;. 
\end{equation} 
Having set 
\mbox{$E=0$}, 
we see that the correct choice of diagonal saddle-point matrix is 
\mbox{$\sigma_0 = -iL$}. 
So we write
\mbox{$\sigma_{\rm G} = -iQ$} 
where 
\mbox{$Q \equiv T^{-1}LT$}. 
The supermatrices $Q$ identically span the coset space of the $T_0$; 
one can replace the coset elements $T_0$ by the group elements $T$ in the 
expression for $Q$, as it is clear that only group elements from different 
cosets give rise to distinct $Q$. 
We now arrive at the large-$N$ form of the ensemble-averaged generating 
function as a zero-dimensional supermatrix non-linear $\sigma$-model, 
\begin{equation} 
\overline{Z(\varepsilon)} = \int {\cal D}Q\, 
     e^{i{\cal L}_{\rm eff}(Q)} 
     e^{i{\cal L}_{\rm source}(Q;J)} \;, 
\label{467}
\end{equation}
where 
\begin{eqnarray}
i{\cal L}_{\rm eff}(Q) & = & -\frac{N\tilde{\omega}}{\lambda}\trg LQ \;, 
\nonumber \\ 
i{\cal L}_{\rm source}(Q;J) & = & \frac{iN}{\lambda} 
     \trg(Q-\lambda^{-1}\tilde{\omega}L)J \;, 
\end{eqnarray} 
and it is convenient to write 
\mbox{${\cal D}Q \equiv {\cal D}\mu(T_0)$}. 
In the present application, we have 
\mbox{$\trg J^2 = 0$}. 
Also, we can neglect $\tilde{\omega}$ in the source Lagrangian because it 
will give rise to contributions that are non-leading in $N^{-1}$. 
Finally, we let 
\mbox{$\omega^+ = \omega + i\epsilon$} 
to obtain 
\mbox{$-N\tilde{\omega}/\lambda = iN\omega^+/(2\lambda) = 
     i\pi\omega^+/(2d)$}. 
Hence we take 
\begin{eqnarray}
i{\cal L}_{\rm eff}(Q) & = & \frac{i\pi\omega^+}{2d}\trg LQ \quad = \quad 
     \frac{i\pi\omega^+}{2d}(\trg Q_{11} - \trg Q_{22}) \;, 
\nonumber \\ 
i{\cal L}_{\rm source}(Q;J) & = & \frac{iN}{\lambda} \trg QJ \quad = \quad 
     \frac{iN}{\lambda}\sum_{p=1,2} \varepsilon_p \trg kQ_{pp} \;. 
\label{469}
\end{eqnarray} 
Therefore, 
\begin{eqnarray} 
N^{-2}\overline{G(E_1^+)G(E_2^-)} & = & \left. \frac{1}{4N^2} 
     \frac{\partial^2}{\partial\varepsilon_1\partial\varepsilon_2} 
     \overline{Z(\varepsilon)}\right|_{\varepsilon=0} 
\nonumber \\ 
& = & -\frac{1}{(2\lambda)^2}\int{\cal D}Q\, \trg kQ_{11} \trg kQ_{22} 
     \exp\left\{\frac{i\pi\omega^+}{2d}\left(\trg Q_{11} - 
     \trg Q_{22}\right)\right\} \;. 
\label{470}
\end{eqnarray} 
It is also useful to note that from Eq.~(\ref{Tww}) it follows 
immediately that 
\mbox{$\trg Q_{22} = -\trg Q_{11}$}. 
% and \mbox{$\trg kQ_{22} = -\trg kQ_{11}$}. 

\chapter{Superintegration} 
The measure 
\mbox{${\cal D}Q$} 
is the Haar measure on the coset space of the supermatrices $Q$. 
If we set 
\mbox{$\varepsilon = 0$}, 
then since by construction 
\mbox{$\overline{Z(0)}=1$} 
and 
\mbox{${\cal L}_{\rm source}(Q;0)=0$}, 
we have 
\begin{equation} 
1 = \int{\cal D}Q\, \exp\Bigl\{\frac{i\pi\omega^+}{2d}\trg LQ\Bigr\} 
\end{equation} 
for any 
\mbox{$\omega^+ \neq 0$}. 
By now taking the limit 
\mbox{$\omega^+ \to 0$}, 
we see that 
\mbox{${\cal D}Q$} 
should be normalized to unit coset volume. 

\section{Coset Parametrization} 
Since 
\begin{equation} 
\left( 
\begin{array}{cc} 
0 & w \\ 
\overline{w} & 0 
\end{array} 
\right) \left(
\begin{array}{cc} 
0 & w \\ 
\overline{w} & 0 
\end{array} 
\right) = \left(
\begin{array}{cc} 
w\overline{w} & 0  \\ 
0 & \overline{w}w 
\end{array} 
\right) \;, 
\end{equation} 
Eq.~(\ref{Tww}) can be expanded to yield 
\begin{equation} 
T_0 = \left( 
\begin{array}{cc} 
\cos\sqrt{w\overline{w}} & iw\,\frac{\sin\sqrt{\overline{w}w}} 
     {\sqrt{\overline{w}w}} \\ 
i\overline{w}\,\frac{\sin\sqrt{w\overline{w}}}{\sqrt{w\overline{w}}} & 
     \cos\sqrt{\overline{w}w} 
\end{array} 
\right) \;. 
\end{equation} 
Let us set 
\begin{equation} 
t_{12} = iw\,\frac{\sin\sqrt{\overline{w}w}}{\sqrt{\overline{w}w}} \;, \quad 
t_{21} = i\overline{w}\,\frac{\sin\sqrt{w\overline{w}}}{\sqrt{w\overline{w}}} 
\end{equation} 
so that 
\mbox{$t_{12}, t_{21}$} 
are $2\times 2$ graded matrices satisfying 
\begin{equation} 
t_{12}^\dagger = kt_{21} \;, \quad 
t_{21}^\dagger = t_{12}k \;. 
\label{t12} 
\end{equation} 
Then we obtain the representation 
\begin{equation} 
T_0 = \left( 
\begin{array}{cc} 
\sqrt{1+t_{12}t_{21}} & t_{12} \\ 
t_{21} & \sqrt{1+t_{21}t_{12}} 
\end{array} 
\right) 
\end{equation} 
for the general coset element $T_0$, which leads to 
\begin{equation} 
Q = \left( 
\begin{array}{cc} 
1+2t_{12}t_{21} & 2t_{12}\sqrt{1+t_{21}t_{12}} \\ 
-2t_{21}\sqrt{1+t_{12}t_{21}} & -(1+2t_{21}t_{12}) 
\end{array} 
\right) \;.
\label{57}
\end{equation} 

Now let us write 
\begin{equation} 
t_{12} = u^{-1}\mu v \;, \quad 
t_{21} = v^{-1}\overline{\mu}u \;, 
\label{udv} 
\end{equation} 
where 
\mbox{$\mu,\overline{\mu}$} 
are diagonal matrices. This serves to diagonalize 
\mbox{$t_{12}t_{21}$} 
and 
\mbox{$t_{21}t_{12}$}, 
\begin{equation} 
t_{12}t_{21} = u^{-1}(\mu\overline{\mu})u \;, \quad 
t_{21}t_{12} = v^{-1}(\overline{\mu}\mu)v \;. 
\label{diag}
\end{equation} 
Such a diagonalization must exist because (i) 
\mbox{$t_{12}t_{21}$} 
is Hermitian and hence diagonalizable, while (ii) 
\mbox{$t_{21}t_{12}$} 
is not Hermitian but has the same eigenvalues as 
\mbox{$t_{12}t_{21}$} 
and so can also be diagonalized. 
Now, from inserting the representations (\ref{diag}) into the 
relations 
\begin{equation} 
(t_{12}t_{21})^\dagger = t_{12}t_{21} \;, \quad 
(t_{21}t_{12})^\dagger = kt_{21}t_{12}k \;, 
\end{equation} 
respectively, and noting that the elements of 
\mbox{$\mu\overline{\mu} = \overline{\mu}\mu$} 
are real, it follows that one can choose 
\begin{equation} 
u^{-1} = u^\dagger \;, \quad 
v^{-1} = kv^\dagger k \;. 
\end{equation} 
Then, for the representation (\ref{udv}) to satisfy Eq.~(\ref{t12}), 
we must have 
\mbox{$\mu^\dagger = k\overline{\mu}$},
so that 
\mbox{$\mu,\overline{\mu}$} 
can be regarded as diagonal special cases of the general matrices 
\mbox{$t_{12},t_{21}$}, 
respectively. 
For $T_0$, Eq.~(\ref{udv}) leads us to the form 
\begin{equation} 
T_0 = U^{-1}T_{\rm d}U \;, \quad 
U = \left( 
\begin{array}{cc} 
u & 0 \\ 
0 & v 
\end{array} 
\right) \;, 
\end{equation} 
where $T_{\rm d}$ is block-diagonal in the graded indices 
\mbox{$\alpha,\alpha'$}, 
\begin{equation} 
T_{\rm d} = \left( 
\begin{array}{cc} 
\sqrt{1+\mu\overline{\mu}} & \mu \\ 
\overline{\mu} & \sqrt{1+\overline{\mu}\mu} 
\end{array} 
\right) \;. 
\label{mu}
\end{equation} 

Thus, let us consider the (bosonic) subgroup of the supergroup of the $T$ that 
is block-diagonal in the graded indices 
\mbox{$\alpha,\alpha'$}. 
Its elements assume the form 
\mbox{$(T_{\rm d})^{\alpha\alpha'}_{pp'} = \delta^{\alpha\alpha'} 
     T^{(\alpha)}_{pp'}$}, 
and we find it convenient to introduce the notation 
\mbox{$T_{\rm d} = \diag(T_{\rm b},T_{\rm f})$}. 
The relation (\ref{sTs}) implies that 
\begin{equation} 
T^{-1}_{\rm b} = LT^\dagger_{\rm b}L \;, \quad 
T^{-1}_{\rm f} = T^\dagger_{\rm f}   \;, 
\end{equation} 
where 
\mbox{$L=\diag(+1,-1)$} 
in the $p$-block indices. 
We see that the subgroup which operates in the boson-boson block is 
\mbox{$T_{\rm b} \in \mbox{U(1,1)}$}, 
while in the fermion-fermion block, 
\mbox{$T_{\rm f} \in \mbox{U(2)}$}. 
However, to generate distinct coset elements 
\mbox{$\Lambda \equiv T^{-1}_{\rm d}LT_{\rm d}$}, 
we should restrict the subgroup elements to range only over the cosets 
\mbox{$T_{\rm b} \in \mbox{SU(1,1)/U(1)}$} (a hyperboloid), 
and 
\mbox{$T_{\rm f} \in \mbox{SU(2)/U(1)}$} (the unit sphere). 

These coset spaces can each be parametrized by two real `angles', according 
to 
\begin{equation} 
T_{\rm b} = \left( 
\begin{array}{cc} 
\cosh \half\theta_0 & -\zeta e^{i\phi_0}\sinh \half\theta_0 \\ 
\zeta^* e^{-i\phi_0}\sinh \half\theta_0 & \cosh \half\theta_0 
\end{array} 
\right) \;, \quad 
T_{\rm f} = \left( 
\begin{array}{cc} 
\cos \half\theta_1 & i\zeta e^{i\phi_1}\sin \half\theta_1 \\ 
i\zeta^* e^{-i\phi_1}\sin \half\theta_1 & \cos \half\theta_1 
\end{array} 
\right) \;,
\end{equation} 
where $\zeta$ is an arbitrary phase, 
\mbox{$|\zeta|=1$}. 
Let us now combine these parameters into $2\times 2$ (super)-matrices given 
by 
\begin{equation} 
\hat{\phi} = \left( 
\begin{array}{rr} 
\phi_0 & 0 \\ 
0      & \phi_1 
\end{array} 
\right) \;, \quad 
\hat{\theta} = \left( 
\begin{array}{rr} 
i\theta_0 & 0 \\ 
0      & \theta_1 
\end{array} 
\right) \;. 
\end{equation} 
Then we can write 
\begin{equation} 
T_{\rm d} = \left(
\begin{array}{cc} 
e^{i\hat{\phi}} & 0 \\ 
0               & {\opfone} 
\end{array} 
\right) \left( 
\begin{array}{cc} 
\cos\half\hat{\theta} & i\zeta\sin\half\hat{\theta} \\ 
i\zeta^*\sin\half\hat{\theta}  & \cos\half\hat{\theta} 
\end{array} 
\right) \left( 
\begin{array}{cc} 
e^{-i\hat{\phi}} & 0 \\ 
0               & {\opfone} 
\end{array} 
\right) \;, 
\end{equation} 
where the displayed block structure now pertains to the $p$-block indices. 

From Eq.~(\ref{mu}), we can make the identifications 
\begin{equation} 
\mu = i\zeta e^{i\hat{\phi}}\sin\half\hat{\theta} \;, 
\quad
\overline{\mu} = i\zeta^* e^{-i\hat{\phi}}\sin\half\hat{\theta} 
\end{equation} 
for the diagonal matrices 
\mbox{$\mu,\overline{\mu}$} 
appearing in Eq.~(\ref{udv}). 
On introducing the `eigenvalues' 
\begin{eqnarray} 
\lambda_0 & = & \cosh\theta_0 \;, \quad \hphantom{-}1 \le \lambda_0 
     < \infty \;, 
\nonumber \\ 
\lambda_1 & = & \cos\theta_1\hphantom{h} \;, \quad -1 \le \lambda_1 \le +1 \;, 
\end{eqnarray} 
we see that 
\begin{eqnarray} 
\mu_0\overline{\mu}_0 & = & \overline{\mu}_0\mu_0 \quad = \quad 
     \half(\lambda_0-1) \;, 
\nonumber \\ 
\mu_1\overline{\mu}_1 & = & \overline{\mu}_1\mu_1 \quad = \quad 
     \half(\lambda_1-1) \;. 
\end{eqnarray} 
Consequently, we have 
\begin{equation} 
e^{i\alpha\trg LQ} = e^{2i\alpha\trg t_{12}t_{21}} = 
     e^{2i\alpha (\lambda_0-\lambda_1)} \;. 
\label{520}
\end{equation} 
Similarly, we obtain 
\begin{eqnarray} 
\trg kQ_{11} & = & \trg \hat{\lambda}uku^{-1} \;, 
\nonumber \\ 
\trg kQ_{22} & = & -\trg \hat{\lambda}vkv^{-1} \;, 
\end{eqnarray} 
where 
\mbox{$\hat{\lambda} \equiv \diag(\lambda_0,\lambda_1)$}. 
We also note that the ensuing representation for $Q$ is then 
\mbox{$Q = U^{-1}\Lambda U$} 
with 
\begin{equation} 
\Lambda = \left( 
\begin{array}{cc} 
e^{i\hat{\phi}} & 0 \\ 
0               & {\opfone} 
\end{array} 
\right) \left( 
\begin{array}{cc} 
\cos\hat{\theta} & i\zeta\sin\hat{\theta} \\ 
-i\zeta^*\sin\hat{\theta}  & -\cos\hat{\theta} 
\end{array} 
\right) \left( 
\begin{array}{cc} 
e^{-i\hat{\phi}} & 0 \\ 
0               & {\opfone} 
\end{array} 
\right) \;. 
\end{equation} 

The coset supermatrices $T_0$ have two complex Grassmannian degrees of 
freedom (say $\eta$ and $\rho$), 
one of which must be contained in $u$, and the other in $v$. These matrices 
have no bosonic degrees of freedom because these are all already accounted 
for in the bosonic subspace spanned by the $T_{\rm d}$. 
The relations 
\mbox{$u^{-1} = u^\dagger$} 
and 
\mbox{$v^{-1} = kv^\dagger k$} 
then imply that we may take 
\begin{eqnarray} 
u & = & \exp\left( 
\begin{array}{rc} 
0 & -\hphantom{i}\eta^* \\ 
\hphantom{i}\eta & 0 
\end{array} 
\right) \quad = \quad \left( 
\begin{array}{cc} 
1-\half\eta^*\eta & -\eta^* \\ 
\eta & 1+\half\eta^*\eta 
\end{array} 
\right) \;, 
\nonumber \\ 
v & = & \exp\left( 
\begin{array}{rc} 
0 & -i\rho^* \\ 
i\rho & 0 
\end{array} 
\right) \quad = \quad \left( 
\begin{array}{cc} 
1+\half\rho^*\rho & -i\rho^* \\ 
i\rho & 1-\half\rho^*\rho 
\end{array} 
\right) \;. 
\end{eqnarray} 
It follows that we also have the Hermiticity properties 
\mbox{$u^\dagger = kuk$} 
and 
\mbox{$v^\dagger = v$}. 
Thus, 
\begin{eqnarray} 
\trg\hat{\lambda}uku^{-1} & = & (\lambda_0+\lambda_1) + 
     2(\lambda_1-\lambda_0)\eta^*\eta \;, 
\nonumber \\ 
\trg\hat{\lambda}vkv^{-1} & = & (\lambda_0+\lambda_1) - 
     2(\lambda_1-\lambda_0)\rho^*\rho \;. 
\label{524}
\end{eqnarray} 

\section{The Measure}
To calculate the measure 
\mbox{${\cal D}Q$} 
in terms of specific manifold coordinates, we first note that the invariant 
line element on the $Q$-manifold is given by 
\begin{eqnarray} 
\frac{1}{16}\trg dQdQ & = & \frac{1}{16}\trg [L, T_0dT_0^{-1}]^2 
\\ & = & \frac{1}{4}\trg\left[ 
     d(t_{12}t_{21})d(t_{12}t_{21}) - 
     d(t_{12}\sqrt{1+t_{21}t_{12}})d(t_{21}\sqrt{1+t_{21}t_{12}}) 
     + (1\leftrightarrow 2) \right] \;. 
\nonumber 
\end{eqnarray} 
Let us now introduce 
\mbox{$\tau_{12}, \tau_{21}$} 
by 
\begin{equation} 
t_{12} = 2(1-\tau_{12}\tau_{21})^{-1}\tau_{12} \;, \quad 
t_{21} = 2(1-\tau_{21}\tau_{12})^{-1}\tau_{21} \;, 
\end{equation} 
to obtain the `rational' parametrization of $T_0$: 
\begin{equation} 
T_0 = \left( 
\begin{array}{cc} 
(1+\tau_{12}\tau_{21})(1-\tau_{12}\tau_{21})^{-1} & 
     2(1-\tau_{12}\tau_{21})^{-1}\tau_{12} \\ 
2(1-\tau_{21}\tau_{12})^{-1}\tau_{21} & 
     (1+\tau_{21}\tau_{12})(1-\tau_{21}\tau_{12})^{-1} 
\end{array} 
\right) \;. 
\end{equation} 
We the aid of this parametrization, one can show that \cite{VWZ} 
\begin{equation} 
{\cal D}Q = \prod_{\alpha,\alpha'} dt_{12}^{\alpha\alpha'} 
     dt_{21}^{\alpha\alpha'} \;, 
\end{equation} 
where it is understood that a complex conjugate, such as 
\mbox{$(t_{12}^{\alpha\alpha'})^*$},
should be expressed in terms of 
\mbox{$t_{21}^{\alpha'\alpha}$}.
It follows from this that 
the result for the invariant integration measure on the GUE coset space 
in terms of eigenvalues and angles is given by 
\begin{equation} 
{\cal D}Q = \frac{1}{(\lambda_0-\lambda_1)^2}d\lambda_0d\lambda_1
     d\phi_0d\phi_1d\eta d\eta^* d\rho d\rho^* \;. 
\end{equation} 
Combining this with Eqs.~(\ref{470}), (\ref{520}) and (\ref{524}), we 
obtain the result 
\begin{eqnarray} 
N^{-2}\overline{G(E^+_1)G(E^-_2)} & = & \frac{1}{4\lambda^2}\int {\cal D}Q\,  
     \trg\hat{\lambda}uku^{-1} \trg\hat{\lambda}vkv^{-1} 
     e^{2i\alpha\trg\hat{\lambda}} 
\nonumber \\ 
& \stackrel{\scriptstyle\rm conn.}{\longrightarrow} & -\frac{1}{\lambda^2} 
     \int {\cal D}Q\, 
     (\lambda_0-\lambda_1)^2 \eta^*\eta \rho^*\rho \, 
     e^{2i\alpha(\lambda_0-\lambda_1)} 
\nonumber \\ 
& = & \frac{1}{\lambda^2}\int_1^\infty d\lambda_0 \int_{-1}^{+1} d\lambda_1\,
     e^{2i\alpha(\lambda_0-\lambda_1)} 
\nonumber \\ 
& = & \frac{2i}{\lambda^2}\left(\frac{d}{\pi\omega^+}\right)^2 
     e^{i\pi\omega^+/d} 
     \sin\left(\frac{\pi\omega^+}{d}\right) \;,  
\label{GGcon}
\end{eqnarray} 
where the abbreviation conn.\ stands for `connected part', and recalling that 
\mbox{$2\alpha = \pi\omega^+/d$}. 
We should point out that the integral over the contribution to 
Eq.~(\ref{GGcon})  
of zeroth order in the Grassmann variables, namely 
\mbox{$(\lambda_0+\lambda_1)^2$}, 
is singular and actually non-vanishing when treated carefully. 
However, this contribution simply 
represents the disconnected part \cite{VZ} of the correlator, which we want 
to subtract in any case. 
To see this, we appeal to a theorem quoted in Ref.~\citenum{Zir} which 
states that the contribution to the super-integral 
\mbox{$\int {\cal D}Q\, f(Q)$} 
from the component of 
\mbox{$f(Q)$} 
that is of zeroth order in the Grassmann variables is simply given by 
\mbox{$f(L)$}, 
i.e.\ the function 
\mbox{$f(Q)$} 
evaluated at the origin of the $Q$-coset manifold. The corresponding 
contribution from the integral in Eq.~(\ref{470}) is thus seen to be 
\begin{eqnarray} 
N^{-2}\overline{G(E^+_1)G(E^-_2)} & \to & -\Bigl(\frac{1}{2\lambda} 
     \trg kL_{11}\Bigr){\cdot}\Bigl(\frac{1}{2\lambda} \trg kL_{22}\Bigr) 
\nonumber \\ 
& = & N^{-2}\overline{G(E^+_1)}\cdot\overline{G(E^-_2)} 
\nonumber \\ 
& = & \lambda^{-2} \;, 
\end{eqnarray} 
i.e., it is just the disconnected part. 
Let us write 
\mbox{$C(\omega) \equiv \left. C(E_1,E_2)\right|_{E=0}$}. 
Then we obtain, from Eq.~(\ref{C12}), the expression 
\begin{equation} 
C(\omega) =  \frac{1}{(\pi\lambda)^2}\biggl[ d\delta(\omega) - 
     \left(\frac{d}{\pi\omega}\right)^2 
     \sin^2\left(\frac{\pi\omega}{d}\right)\biggr]  
\end{equation} 
for the connected part of the density-density correlator at the centre of 
the spectrum. 
The $\delta$-function arises from the singular contribution to the real part of
Eq.~(\ref{GGcon}) when 
\mbox{$\omega \to 0$}, 
which is induced by the positive imaginary part of 
\mbox{$\omega^+ = \omega + i\epsilon$}. 
The corresponding term is given by 
\begin{equation}
\frac{2d}{\pi\lambda^2}\re\left(\frac{i}{\omega^+}\right) \quad = \quad 
     \frac{2d}{\lambda^2}\frac{1}{2\pi i}\left(\frac{1}{\omega-i\epsilon}
     - \frac{1}{\omega+i\epsilon}\right) \quad \asym{\epsilon \to 0^+} \quad 
     \frac{2d}{\lambda^2}\delta(\omega)\;.
\end{equation}

A description of the modifications required of the present formalism in 
order to deal with the GOE is given in Ref.~\citenum{PWZL2}. 
The presence of time-reversal symmetry (manifested in the real-symmetric 
nature of the Hamiltonian $H$) necessitates a doubling of the dimension of 
the supervector $\varphi$ before performing the Hubbard-Stratonovich 
transformation, viz.\ 
\mbox{$\Phi = (\varphi, s\varphi^*)^{\rm T}$}, 
so that a Majorana-type condition 
\mbox{$\Phi^* = C\Phi$} 
is satisfied, where $C$ is an appropriate `charge-conjugation' matrix. 
This entails a doubling of the dimension of the graded matrices $Q$, 
supplemented by a reality condition 
\mbox{$Q^* = CQC^{-1}$}. 

\section*{Acknowledgements} 
I thank Prof.~H.A.~Weidenm\"uller and the members of the Theory Group at the 
Max-Planck-Institut f\"ur Kernphysik in Heidelberg for hospitality and 
financial support during a visit when this work was carried out. I am 
indebted to the postdocs and students, especially Ulrich Gerland, who 
worked through the manuscript.
Financial support from NSERC (Canada) in the form of a research grant 
is also gratefully acknowledged. 

\appendix
\chapter{} 
\renewcommand{\arraystretch}{0.5} 
In this appendix, we collect the various constant matrices that appear
throughout the paper and we present derivations of the matrix Gaussian
integrals over commuting and anti-commuting variables that appear in
Eqs.~(\ref{bosonic}) and (\ref{fermionic}), respectively. 
Matrices arising from the graded structure of the theory and the
symmetry-breaking due to the presence of advanced and retarded components are
given by 
\begin{equation}
k^{\alpha\alpha'} = \left(
\begin{array}{rr}
1 &  0 \\ 
0 & -1 
\end{array}
\right)_{\alpha\alpha'} \;, \quad 
L_{pp'} = \left(
\begin{array}{rr}
{\opfone} &  0 \\ 
0 & -{\opfone} 
\end{array}
\right)_{pp'} \;, \quad
s_{pp'} = \left(
\begin{array}{rr}
{\opfone} &  0 \\ 
0        & -k
\end{array}
\right)_{pp'} \;. 
\end{equation}
Matrices associated with the source terms are
given by 
\begin{equation}
I_{pp'}(1) = \left(
\begin{array}{rr}
 1 & 0 \\ 
 0 & 0 
\end{array}
\right)_{pp'} \;, \quad 
I_{pp'}(2) = \left(
\begin{array}{rr}
 0 & 0 \\ 
 0 & 1 
\end{array}
\right)_{pp'} \;. 
\end{equation}
In all cases, the explicit indices indicate the space to which the displayed 
block structure pertains. 

To derive Eq.~(\ref{bosonic}), we first observe that, since the matrix $A$
that appears in the exponent is assumed to be positive Hermitian, it can
be diagonalized according to 
\mbox{$A = U^\dagger A_{\rm D}U$}, 
where $U$ is a unitary matrix and $A_{\rm D}$ is the diagonal matrix of
the positive eigenvalues, 
\mbox{$A_{\rm D} = \diag(\lambda_1,\lambda_2,\ldots,\lambda_N)$}. 
This decomposition facilitates the linear transformation of integration
variables 
\begin{equation} 
z \mapsto z' = Uz \;, \quad 
z^\dagger \mapsto {z'}^\dagger = z^\dagger U^\dagger \;, 
\end{equation} 
whose Jacobian is equal to unity. The resulting decoupling of the
integration variables $z'$ gives rise to a factorized form of the Gaussian
integral in Eq.~(\ref{bosonic}): 
\begin{equation} 
I_0(A) = \prod_{k=1}^N\frac{1}{2\pi i}\int dz'_k \int d{z'_k}^*\, 
     e^{-\lambda_k {z'_k}^*z'_k} \;. 
\end{equation} 
Next, we pass to the real and imaginary parts of the holomrphic variables
\mbox{$z'_k = x_k+iy_k$}, 
\mbox{${z'_k}^* = x_k-iy_k$}, 
in which case 
\begin{eqnarray} 
I_0(A) & = & \prod_{k=1}^N\frac{1}{\pi}\int_{-\infty}^{+\infty}dx_k\, 
     e^{-\lambda_k x^2_k} \int_{-\infty}^{+\infty} dy_k\, 
     e^{-\lambda_k y^2_k}
\nonumber \\
& = & \prod_{k=1}^N 1/\lambda_k 
\nonumber \\
& = & (\det A)^{-1} \;, 
\end{eqnarray} 
having appealed to the fact that the product of eigenvalues coincides with
the determinant. 

The analogous integral over anti-commuting variables Eq.~(\ref{fermionic})
is most easily derived by expanding the integrand in a power series. Since
\mbox{$\eta^2_\ell = \eta_\ell^{*2} =0$} 
and we have $2N$ Grassmann degrees of freedom 
(\mbox{$\eta_\ell, \eta^*_\ell$} 
for 
\mbox{$\ell = 1,2,\ldots,N$}), 
only terms up to order $2N$ can survive this power-series expansion of the
exponential. On the other hand, in any of the ensuing polynomial terms
that are of order less than $2N$, at least one Grassmann degree of freedom
(say $\eta_k$) will be missing. 
Since, by Eq.~(\ref{zero}), 
\mbox{$\int d\eta_k =0$}, 
it follows that such terms vanish on integration. Thus, only the
polynomial of order $2N$ survives; and so we can write for the Gaussian
integral in Eq.~(\ref{fermionic}) 
\begin{equation} 
I_1(A) = \int\prod_{\ell=1}^N \frac{d\eta^*_\ell d\eta_\ell}{2\pi}\, 
     \frac{(-1)^N}{N!} (\eta^\dagger A\eta)^N \;. 
\label{star2} 
\end{equation} 
Now, 
\begin{equation} 
\eta^\dagger A\eta = \sum_{i,j=1}^N \eta^*_i A_{ij}\eta_j \;. 
\end{equation} 
Again, since 
\mbox{$\eta^2_\ell = \eta_\ell^{*2} =0$}, 
we have 
\begin{eqnarray} 
(\eta^\dagger A\eta)^N & = & \sum_{\sigma,\tau\in {\rm P}_N} 
     \prod_{\ell=1}^N \eta^*_{\sigma(\ell)}A_{\sigma(\ell),\tau(\ell)} 
     \eta_{\sigma(\ell)} 
\nonumber \\ 
& = &  \sum_{\sigma,\tau\in {\rm P}_N}\biggl(\prod_{\ell=1}^N 
     A_{\sigma(\ell),\tau(\ell)}\biggr)\biggl(\prod_{\ell=1}^N 
     \eta^*_{\sigma(\ell)} \eta_{\tau(\ell)}\biggr) 
\end{eqnarray} 
so that each $\eta_\ell$ and $\eta^*_\ell$ appears exactly once for all 
\mbox{$\ell = 1,2,\ldots,N$}, 
and where 
\mbox{$\sigma,\tau$} 
range over the group of permutations of $N$ objects ${\rm P}_N$. 

After some elementary manipulation of the product of Grassmann variables
above to achieve the relation 
\begin{equation} 
\prod_{\ell=1}^N \eta^*_{\sigma(\ell)}\eta_{\tau(\ell)} \quad = \quad 
     (-1)^\sigma\prod_{\ell=1}^N \eta^*_\ell\eta_{\tau(\ell)} \quad=\quad 
     (-1)^\sigma(-1)^\tau\prod_{\ell=1}^N \eta^*_\ell\eta_\ell \;, 
\end{equation} 
where $(-1)^\sigma$ equals $+1$ or $-1$ if the permutation $\sigma$ is
even or odd, respectively, we can recast the integral (\ref{star2}) into
the form 
\begin{eqnarray} 
I_1(A) & = & \frac{1}{N!}\sum_{\sigma,\tau\in {\rm P}_N}(-1)^\sigma 
     (-1)^\tau \prod_{\ell=1}^N A_{\sigma(\ell),\tau(\ell)} {\cdot} 
     \prod_{\ell=1}^N \int\frac{d\eta_\ell d\eta^*_\ell}{2\pi}\, 
     \eta^*_\ell\eta_\ell 
\nonumber \\ 
& = & \frac{1}{N!}\sum_{\sigma,\tau\in {\rm P}_N}(-1)^\sigma 
     (-1)^\tau \prod_{\ell=1}^N A_{\sigma(\ell),\tau(\ell)} \;, 
\label{hash2} 
\end{eqnarray} 
having made use of Eq.~(\ref{45}). 
Next, let us write 
\mbox{$\ell = \sigma^{-1}(k)$} 
for appropriate 
\mbox{$k = 1,2,\ldots,N$}. 
Then 
\begin{equation} 
\prod_{\ell=1}^N A_{\sigma(\ell),\tau(\ell)} \quad=\quad 
     \prod_{k=1}^N A_{\sigma(\sigma^{-1}(k)),\tau(\sigma^{-1}(k))} 
     \quad=\quad \prod_{k=1}^N A_{k,\tau\circ\sigma^{-1}(k)} \;. 
\end{equation} 
Hence, 
\begin{eqnarray} 
I_1(A) & = & \frac{1}{N!}\sum_{\sigma\in {\rm P}_N}
     \sum_{\tau\in {\rm P}_N} (-1)^{\tau\circ\sigma^{-1}} 
     \prod_{k=1}^N A_{k,\tau\circ\sigma^{-1}(k)} 
\nonumber \\ 
& = & \frac{1}{N!}\sum_{\sigma,\nu\in {\rm P}_N} (-1)^\nu 
     \prod_{k=1}^N A_{k,\nu(k)} 
\nonumber \\ 
& = & \sum_{\nu\in {\rm P}_N} (-1)^\nu \prod_{k=1}^N A_{k,\nu(k)} 
\nonumber \\ 
& = & \sum_{\nu\in {\rm P}_N} (-1)^\nu A_{1,\nu(1)}A_{2,\nu(2)} 
     \cdots A_{N,\nu(N)} 
\nonumber \\ 
& = & \det A 
\end{eqnarray} 
by the definition of the determinant. 

\chapter{} 
If we write for the block-diagonal matrix $R$ introduced above 
Eq.~(\ref{TPT0}), 
\mbox{$R = \diag(R_1,R_2)$}, 
then 
\linebreak 
\mbox{$R^{-1} = sR^\dagger s$} 
implies that 
\mbox{$R_1^{-1} = R_1^\dagger$} 
and 
\mbox{$R_2^{-1} = kR_2^\dagger k$}. 
Since $R_1$ is a unitary $2\times 2$ supermatrix, we can parametrize it as 
\begin{equation} 
R_1 = \exp\biggl\{ i \left( 
\begin{array}{cc} 
\chi^{\rm b}_1 & 0 \\ 
0 & \chi^{\rm f}_1 
\end{array} 
\right)\biggr\} \exp \left( 
\begin{array}{cc} 
0 & -\xi^*_1 \\ 
\xi_1 & 0 
\end{array} 
\right) \;. 
\end{equation} 
Then, since 
\mbox{$\delta P = R^{-1}P_{\rm d}R$}, 
we have 
\begin{eqnarray} 
\delta P_1 & = & \left( 
\begin{array}{cc} 
1-\half\xi_1^*\xi_1 & \xi_1^* \\ 
-\xi_1 & 1+\half\xi_1^*\xi_1 
\end{array} 
\right) \left( 
\begin{array}{cc} 
a_1 & 0 \\ 
0 & ib_1 
\end{array} 
\right) \left( 
\begin{array}{cc} 
1-\half\xi_1^*\xi_1 & -\xi_1^* \\ 
\xi_1 & 1+\half\xi_1^*\xi_1 
\end{array} 
\right) 
\nonumber \\ 
& = & 
\left( 
\begin{array}{cc} 
a_1 -(a_1-ib_1)\xi_1^*\xi_1 & (a_1-ib_1)\xi^*_1 \\ 
(a_1-ib_1)\xi_1 & ib_1 -(a_1-ib_1)\xi_1^*\xi_1 
\end{array} 
\right) \;. 
\end{eqnarray} 
Since $R_2$ is a pseudo-unitary $2\times 2$ supermatrix, 
we can parametrize it as 
\begin{equation} 
R_2 = \exp\biggl\{ i\left( 
\begin{array}{cc} 
\chi^{\rm b}_2 & 0 \\ 
0 & \chi^{\rm f}_2 
\end{array} 
\right)\biggr\} \exp \biggl\{ i\left( 
\begin{array}{cc} 
0 & -\xi^*_2 \\ 
\xi_2 & 0 
\end{array} 
\right)\biggr\} \;. 
\end{equation} 
It follows that 
\begin{equation} 
\delta P_2 = \left( 
\begin{array}{cc} 
a_2 +(a_2-ib_2)\xi_2^*\xi_2 & (b_2+ia_2)\xi^*_2 \\ 
(b_2+ia_2)\xi_2 & ib_2 +(a_2-ib_2)\xi_2^*\xi_2 
\end{array} 
\right) \;. 
\end{equation} 

The nilpotent parts on the diagonal can be eliminated by appropriate `contour 
deformations'. Then in both cases 
\mbox{$p = 1,2$}, 
we have the form
\begin{equation} 
\delta P_p = \left( 
\begin{array}{cc} 
x & \bar{\eta} \\ 
\eta & iy 
\end{array} 
\right) \;, \quad 
-\infty < x,y < +\infty \;, 
\end{equation} 
with 
\begin{eqnarray} 
{\cal D}P_p & = & -idP^{00}_pdP^{01}_pdP^{10}_pdP^{11}_p 
\nonumber \\ 
& = & dx\,dy\,d\bar{\eta}\,d\eta 
\nonumber \\ 
& = & -rdr\,d\theta\,d\eta\,d\bar{\eta} \;.  
\end{eqnarray} 
Thus, 
\mbox{$\trg(\delta P_p)^2 = r^2 + 2\bar{\eta}\eta$}, 
and so, from Eq.~(\ref{465}), 
\begin{equation} 
K_p  =  -\int_0^{2\pi}d\theta \int_0^\infty rdr\, e^{-c_pr^2/2} 
     \int d\eta\, d\bar{\eta}\, e^{-c_p\bar{\eta}\eta} 
= 1 \;, 
\end{equation} 
recalling the normalization of Grassmann integrals in Eq.~(\ref{45}). 

\chapter{} 
A direct representation of the density of states can be obtained by taking 
for the source matrix 
\linebreak 
\mbox{$J = \varepsilon M = \varepsilon k \otimes L$}. 
Then, from Eq.~(\ref{genfn}), 
\begin{eqnarray} 
\left. \frac{\partial}{\partial\varepsilon} 
     Z(\varepsilon)\right|_{\varepsilon=0} & = & -\Trg kLD^{-1} 
\nonumber \\ 
& = & -2\left[\Tr G^+(E) -\Tr G^-(E)\right] 
\nonumber \\ 
& \stackrel{\epsilon\to 0}{\longrightarrow} & 4i\pi N\rho(E) \;, 
\end{eqnarray} 
noting that here we should take 
\mbox{$\tilde{\omega} = \epsilon$}. 
From Eqs.~(\ref{467}) and (\ref{469}), we have 
\begin{equation} 
\left. \frac{\partial}{\partial\epsilon}
     \overline{Z(\varepsilon)}\right|_{\varepsilon=0} = 
     \frac{iN}{\lambda}\int {\cal D}Q\, \trg kLQ \, 
     \exp \Bigl\{ -\frac{\pi\epsilon}{2d}\trg LQ \Bigr\}\;, 
\end{equation} 
which implies 
\begin{equation} 
\overline{\rho(0)} = \frac{1}{4\pi\lambda} \lim_{\epsilon\to 0^+} 
     \int {\cal D}Q\, \trg kLQ \, 
     \exp \Bigl\{ -\frac{\pi\epsilon}{2d}\trg LQ \Bigr\}\;. 
\end{equation} 
Clearly, only the part of the integrand that is of zeroth order in the 
Grassmann variables contributes. Thus, by the theorem cited below 
Eq.~(\ref{GGcon}), we obtain 
\begin{equation} 
\overline{\rho(0)} = \frac{1}{4\pi\lambda} \trg kL^2 = 
     \frac{1}{\pi\lambda} \;. 
\end{equation} 
Also, from Eq.~(\ref{gsuper}) and (\ref{L1}), 
\begin{eqnarray} 
\left. \frac{\partial}{\partial\epsilon} 
     \overline{Z(\varepsilon)}\right|_{\varepsilon=0} & = & 
     \int {\cal D}\varphi {\cal D}\overline{\varphi}\, 
     \overline{e^{i{\cal L}_1(\varphi;0)}}\, 
     i\trg kL\Bigl(\sum_\mu \varphi(\mu)\overline{\varphi}(\mu)\Bigr) 
\nonumber \\ 
& = & \int {\cal D}\varphi {\cal D}\overline{\varphi}\, 
     e^{i{\cal L}_2(S;0)}\, i\trg kLS \;. 
\end{eqnarray} 
Thus, 
\begin{equation} 
{<}\trg kLS {>}_2 \quad = \quad \frac{N}{\lambda} {<}\trg kLQ{>}_3 
     \quad = \quad 4\pi N\overline{\rho} \;. 
\end{equation} 
This establishes the relation between the expectation of $Q$ in the 
${\cal L}_3$-theory and $S$ in the ${\cal L}_2$-theory. 
It is clear that 
\mbox{${<}Q{>}_3 = L$}. 

Another expression for the average density of states can be obtained by 
appealing to the so-called Ward identity, 
\begin{eqnarray} 
G^+(E) - G^-(E) & = & \Tr\frac{1}{E-H+i\epsilon} - \Tr\frac{1}{E-H-i\epsilon} 
\nonumber \\ 
& = & -2i\epsilon \Tr \frac{1}{(E-H)^2+\epsilon^2} \;, 
\label{res2}
\end{eqnarray} 
in which case 
\begin{eqnarray} 
\rho(E) & = & \frac{1}{\pi N}\lim_{\epsilon\to 0^+}\epsilon 
     \Tr\frac{1}{E-H+i\epsilon}\cdot\frac{1}{E-H-i\epsilon}
\nonumber \\ 
& = & \frac{1}{\pi N}\lim_{\epsilon\to 0^+}\epsilon 
     \Tr D^{-1}_{11} D^{-1}_{22} \;. 
\end{eqnarray} 
Here, we simply take 
\mbox{$\tilde{\omega} = \epsilon$}. 
A source matrix that will generate this combination is furnished by 
Eq.~(\ref{412}) with 
\mbox{$M_m = k_m\otimes {\cal M}$}, 
\mbox{$m = B,F$}, 
where 
\begin{equation} 
{\cal M}_{pp'} = \left( 
\begin{array}{cc} 
0 & 1 \\ 
1 & 0 
\end{array} 
\right)_{pp'} \;, \quad 
k_B^{\alpha\alpha'} = \left( 
\begin{array}{cc} 
1 & 0 \\ 
0 & 0 
\end{array} 
\right)_{\alpha\alpha'} \;, \quad 
k_F^{\alpha\alpha'} = \left( 
\begin{array}{cc} 
0 & 0 \\ 
0 & 1 
\end{array} 
\right)_{\alpha\alpha'} \;.
\end{equation} 
Then, 
\begin{equation}
\left. \frac{\partial^2}{\partial\varepsilon_B\partial\varepsilon_B} 
     Z(\varepsilon)\right|_{\varepsilon=0} = 
     2\Trg D^{-1}_{11}D^{-1}_{22} \;. 
\end{equation}
A similar source matrix was used in Ref.~\citenum{IWZ} for calculating 
moments of conductance. 
In the present case, we have 
\mbox{$\trg LJ = 0$} 
but 
\mbox{$\trg J^2 = 2\trg (\varepsilon{\cdot}k)^2 = 2(\varepsilon^2_B - 
     \varepsilon^2_F)$}. 
Also, 
\begin{eqnarray} 
\trg QJ & = & \trg(\varepsilon{\cdot}k)Q_{12} + 
     \trg(\varepsilon{\cdot}k)Q_{12} 
\nonumber \\ 
& = & \varepsilon_B \trg k_B(Q_{12}+Q_{21}) + 
     \varepsilon_F \trg k_F(Q_{12}+Q_{21}) \;.  
\end{eqnarray} 
It follows that 
\begin{equation} 
\left.\frac{\partial^2}{\partial\varepsilon^2_B} 
     e^{i{\cal L}_{\rm source}(Q;J)}\right|_{\varepsilon=0} = 
     -\frac{N^2}{\lambda^2}\left[\trg k_B(Q_{12}+Q_{21})\right]^2 \;, 
\end{equation} 
whence 
\begin{equation} 
\overline{\rho(0)} = -\frac{N}{2\pi\lambda^2}\lim_{\epsilon\to 0^+} 
     \epsilon \int {\cal D}Q \left[\trg k_B(Q_{12}+Q_{21})\right]^2 
     \exp\Bigl\{ -\frac{\pi\epsilon}{2d}\trg LQ\Bigr\} \;. 
\end{equation} 
We see immediately that only the term in the integrand of highest order in 
Grassmann variables contributes to this expression. 
Since, from Eqs.~(\ref{57}) and (\ref{udv}), 
\begin{equation} 
Q_{12} = 2u^{-1}\mu\sqrt{1+\overline{\mu}\mu}\, v \;, \quad 
Q_{21} = 2v^{-1}\overline{\mu}\sqrt{1+\mu\overline{\mu}}\, u \;, 
\end{equation} 
we have 
\begin{eqnarray} 
\trg k_B(Q_{12}+Q_{21}) & = & 2\bigl[(\mu_0+\overline{\mu}_0) 
     (1+\mu_0\overline{\mu}_0)(1-\half\eta^*\eta)(1+\half\rho^*\rho) 
\nonumber \\ 
& &  {}-i\sqrt{1+\mu_1\overline{\mu}_1}(\mu_1\rho\eta^* + 
     \overline{\mu}_1\eta\rho^*)\bigr] \;. 
\end{eqnarray} 
Thus, the contributing part of the integrand is given by 
\begin{eqnarray} 
\left[\trg k_B(Q_{12}+Q_{21})\right]^2 
     & \stackrel{\scriptstyle\rm c.p.}{\longrightarrow} & -8\bigl[ 
     (1+\mu_0\overline{\mu}_0)\mu_0\overline{\mu}_0 -
     (1+\mu_1\overline{\mu}_1)\mu_1\overline{\mu}_1 
\nonumber \\ 
& &  {}+\half(1+\mu_0\overline{\mu}_0)(\mu_0^2+\overline{\mu}_0^2) 
     \bigr]\eta^*\eta\rho^*\rho \;. 
\end{eqnarray} 
Now, 
\begin{equation} 
(1+\mu_0\overline{\mu}_0)\mu_0\overline{\mu}_0 -
     (1+\mu_1\overline{\mu}_1)\mu_1\overline{\mu}_1 = 
     {\textstyle \frac{1}{4}}(\lambda_0^2-\lambda_1^2) \;, 
\end{equation} 
while the third term in the square brackets above produces a vanishing 
contribution after integration over the angles 
\mbox{$\phi_0,\phi_1$}. 
Hence, we obtain 
\begin{equation} 
\overline{\rho(0)} = \frac{1}{2\pi\lambda}\lim_{\epsilon'\to 0^+}\epsilon' 
     \int_1^\infty d\lambda_0 \int_{-1}^{+1} d\lambda_1\, 
     \frac{\lambda_0+\lambda_1}{\lambda_0-\lambda_1} 
     e^{-\epsilon'(\lambda_0-\lambda_1)} \;, 
\label{avrho}
\end{equation} 
on recalling that 
\mbox{$d = \pi\lambda/N$} 
and setting 
\mbox{$\epsilon' = 2\pi\epsilon/d$}. 
We can write the integral in Eq.~(\ref{avrho}) as 
\begin{equation} 
{\cal I} \equiv \int_{\epsilon'}^\infty dz \int_1^\infty d\lambda_0 
     \int_{-1}^{+1} d\lambda_1\, (\lambda_0+\lambda_1) 
     e^{-z(\lambda_0-\lambda_1)} \;. 
\end{equation} 
The variable transformation 
\begin{equation} 
\lambda_0 = 1+2t_0 \;, \quad \lambda_1=1-2t_1 
\end{equation} 
yields the form 
\begin{eqnarray} 
{\cal I} & = & 8\int_{\epsilon'}^\infty dz \int_0^1 dt_1\, e^{-2zt_1} 
     \int_0^\infty dt_0\, [(1-t_1)+t_0] e^{-2zt_0} 
\nonumber \\ 
& = & 2\int_{\epsilon'}^\infty \frac{dz}{z^2}
\nonumber \\ 
& = & \frac{2}{\epsilon'} \;. 
\end{eqnarray} 
This leads immediately to the expected result 
\mbox{$\overline{\rho(0)} = 1/(\pi\lambda)$}. 

Let us explain why we refer to Eq.~(\ref{res2}) as a Ward identity. 
With a slight modification of our notation, we have from Eqs.~(\ref{467}) 
and (\ref{469}), 
\begin{equation} 
\overline{Z(J+i\tilde{\omega}L)} = \int {\cal D}Q\, \exp\Bigl\{ 
     \frac{iN}{\lambda}\trg Q(J+i\tilde{\omega}L)\Bigr\} \;. 
\end{equation} 
Thus, we can also write 
\begin{equation} 
\overline{Z(A)} = \int {\cal D}Q\, \exp\Bigl\{ 
     \frac{iN}{\lambda}\trg QA\Bigr\} 
\end{equation} 
for any supermatrix $A$. Now let 
\mbox{$A=A_0 + \delta A$} 
where $\delta A$ is infinitesimal. Then we have 
\begin{equation} 
\overline{Z(A_0+\delta A)} - \overline{Z(A_0)} = \frac{iN}{\lambda} 
     \int {\cal D}Q\, 
     \exp\Bigl\{\frac{iN}{\lambda}\trg QA_0\Bigr\}\, 
     \trg Q\delta A \;. 
\end{equation} 
Our theory has the symmetry property 
\begin{equation} 
\overline{Z(T^{-1}(J+i\tilde{\omega}L)T)} =  \overline{Z(J+i\tilde{\omega}L)}
     \;, 
\end{equation} 
which is evident both from Eq.~(\ref{gsuper}) and Eq.~(\ref{467}). 
For infinitesimal transformations, 
\mbox{$T={\opfone}+\delta T$}, 
\mbox{$T^{-1}={\opfone}-\delta T$}, 
this reads 
\begin{equation} 
\overline{Z(J+i\tilde{\omega}L + [J+i\tilde{\omega}L, \delta T])} 
     -  \overline{Z(J+i\tilde{\omega}L)} = 0\;. 
\end{equation} 
So if we choose 
\mbox{$A_0 = J+i\tilde{\omega}L$}, 
\mbox{$\delta A = [A_0,\delta T]$}, 
then we obtain the Ward identity 
\begin{equation} 
\int {\cal D}Q\, e^{{\textstyle \frac{iN}{\lambda}}\trg QA_0}\, 
     [A_0,Q] = 0 \;, 
\end{equation} 
noting that 
\begin{equation} 
\trg Q\delta A = -\trg \delta T[A_0,Q] 
\end{equation} 
and $\delta T$ is arbitrary. 
Equivalently, we can write 
\begin{equation} 
\int {\cal D}Q\, [J,Q]\, 
     e^{{\textstyle \frac{iN}{\lambda}}\trg Q(J+i\tilde{\omega}L)}\, 
     = -i\tilde{\omega} \int {\cal D}Q\, [L,Q]\, 
     e^{{\textstyle \frac{iN}{\lambda}}\trg Q(J+i\tilde{\omega}L)} \;. 
\label{Ward}
\end{equation} 

We can use Eq.~(\ref{Ward}) to generate families of integral identities. 
For example, since on the RHS
\begin{equation} 
[L,Q] = 2\left( 
\begin{array}{cc} 
0       & Q_{12} \\ 
-Q_{21} & 0 
\end{array} 
\right) \;, 
\end{equation} 
we have 
\begin{eqnarray} 
\trg k_\alpha{\cal M}L[L,Q] & = & 2\trg k_\alpha (Q_{12}+Q_{21}) \;, 
\nonumber \\ 
\trg k_\alpha{\cal M}L[J,Q] & = & 2\trg [k_\alpha{\cal M}L,J]Q 
\nonumber \\ 
& = & 2\trg\left( 
\begin{array}{cc} 
-(k_\alpha J_{12}+J_{21}k_\alpha) & -k_\alpha J_{22}+J_{11}k_\alpha \\ 
k_\alpha J_{11} - J_{22}k_\alpha  & k_\alpha J_{12} + J_{21}k_\alpha 
\end{array} 
\right)Q \;, 
\end{eqnarray} 
for 
\mbox{$\alpha = B,F$}, etc.
If we suppose that 
\mbox{$[k_\alpha,J_{pp'}]=0$}, 
then  
\begin{equation} 
\trg k_\alpha{\cal M}L[J,Q] = -\trg k_\alpha(J_{12}+J_{21})LQ 
     +\trg k_\alpha(J_{11}-J_{22}){\cal M}Q \;, 
\end{equation} 
and the Ward identity (\ref{Ward}) now reads 
\begin{eqnarray} 
\int {\cal D}Q\, \left[\trg k_\alpha(J_{11}-J_{22})(Q_{12}+Q_{21}) - 
     \trg k_\alpha(J_{12}+J_{21})(Q_{11}-Q_{22})\right]\,
     e^{{\textstyle \frac{iN}{\lambda}}\trg Q(J+i\tilde{\omega}L)}
& & \nonumber \\ 
=  \quad -2i\tilde{\omega} \int {\cal D}Q\, \trg k_\alpha(Q_{12}+Q_{21})\,
     e^{{\textstyle \frac{iN}{\lambda}}\trg Q(J+i\tilde{\omega}L)} \;. 
\end{eqnarray} 
Let us also take 
\mbox{$J = \varepsilon_\beta k_\beta L + 
     \varepsilon_\gamma' k_\gamma {\cal M}$}, 
so that 
\begin{equation} 
J_{11}-J_{22} = 2\varepsilon_\beta k_\beta \;, \quad 
J_{12}+J_{21} = 2\varepsilon_\gamma' k_\gamma \;, 
\end{equation} 
while 
\begin{equation} 
\trg QJ = \varepsilon_\beta\trg k_\beta(Q_{11}-Q_{22}) + 
     \varepsilon_\gamma'\trg k_\gamma(Q_{12}+Q_{21}) \;. 
\end{equation} 
Then setting 
\mbox{$\alpha=\beta=\gamma$} 
and equating powers of $\varepsilon$ and $\varepsilon'$ yields the family of 
relations 
\begin{equation} 
F(m+1,n+1) = \frac{\lambda}{N\tilde{\omega}}\Bigl[(m+1)F(m,n+1) 
     - nF(m+2,n-1)\Bigr] \;, 
\end{equation} 
valid for 
\mbox{$n = 0,1,2,\ldots$} 
and 
\mbox{$m = -1,0,1,2,\ldots$}, 
where 
\begin{equation} 
F(m,n) = \int{\cal D}Q\, e^{-{\textstyle \frac{N\tilde{\omega}}{\lambda}} 
     \trg QL}\, \left[\trg k_\alpha(Q_{11}-Q_{22})\right]^m 
     \left[\trg k_\alpha(Q_{12}+Q_{21})\right]^n \;. 
\end{equation} 
This constitutes the basis for a recurrence relation for the $Q$-integrals 
\mbox{$F(m,n)$}. 
The case 
\mbox{$m=-1$}, 
\mbox{$n=1$} 
yields an identity equivalent to Eq.~(\ref{res2}). 
We see that the Ward identity relates the transverse modes 
\mbox{$Q_{12},Q_{21}$} 
to the longitudinal modes 
\mbox{$Q_{11},Q_{22}$}. 
\renewcommand{\arraystretch}{1.0} 
\end{document}